\documentclass[12pt,a4paper]{article}

\usepackage{amssymb}
\usepackage[T1]{fontenc}
\usepackage{graphicx}
\usepackage{slashed}
\usepackage{accents}
\usepackage{bbm}
\usepackage{hyperref}

\DeclareMathAlphabet{\boldmathe}{T1}{cmr}{bx}{it}
\newcommand{\mbfgr}[1]{\textit{\mbox{\boldmath$#1$}}}
\newlength{\shiftfeyn}
\newlength{\shiftsubscript}
\def\mbvarphi{\mbfgr{\varphi}}
\def\mbxi{\mbfgr{\xi}}

\def\pa{\partial}

\def\Wb{\bar{W}}
\def\Wt{\tilde{W}}

\def\half{\frac{1}{2}}

\def\pah{\ring {\partial}}
\def\pab{\partial^{\rm b}}

\def\paslac{\partial^{\:\!\rm slac}}
\def\mtxt#1{\quad\hbox{{#1}}\quad}
\newcommand{\ft}[2]{{\textstyle\frac{#1}{#2}}}

\newcommand{\eqnl}[2]{\begin{eqnarray}#1\label{#2}\end{eqnarray}}
\newcommand{\eqngrl}[3]{
\begin{eqnarray}#1\nonumber\\#2\label{#3}\end{eqnarray}}

\newcommand{\refs}[1]{(\ref{#1})}
\newcommand{\de}{\delta}
\renewcommand{\l}{\lambda}
\newcommand{\phb}{\bar{\phi}}
\newcommand{\psb}{\bar{\psi}}
\newcommand{\ve}{\varepsilon}
\newcommand{\veb}{\bar{\ve}}

\newcommand{\one}{\mathbbm{1}}

\newcommand{\TheRevision}{}
\newcommand{\Revision}[1]{\renewcommand{\TheRevision}{#1}}
\begin{document}
\begin{titlepage}
\par
\vskip .5 truecm
\centerline{\large\textbf{ 
Low-dimensional Supersymmetric Lattice Models\footnote{Supported
by the Deutsche Forschungsgemeinschaft, DFG-Wi 777/8-2}}}
\par
\vskip 1 truecm
\begin{center}
\textbf{G.~Bergner, T.~Kaestner, S.~Uhlmann and A.~Wipf}\footnote{e--mail:
G.Bergner, T.Kaestner, A.Wipf@tpi.uni-jena.de and S.Uhlmann@uni-jena.de}\\[2mm]
\it{Theoretisch-Physikalisches Institut, \\
Friedrich-Schiller-Universit\"at Jena\\ 
Fr\"obelstieg 1, D-07743 Jena, Germany}\\
\vskip .5truecm
\end{center}
\par
\vskip 2 truecm

\hypersetup{%colorlinks,linkcolor=blue,citecolor=red,urlcolor=blue,
bookmarksopen,%
pdftitle={Low-dimensional Supersymmetric Lattice Models},pdfauthor={G.~Bergner,
T.~Kaestner, S.~Uhlmann and A.~Wipf (FSU Jena)}}

\begin{abstract}

\noindent
We study and simulate ${\cal N}=2$ supersymmetric Wess-Zumino models in one
and two dimensions. For any choice of the lattice derivative, the theories
can be made manifestly supersymmetric by adding appropriate improvement terms
corresponding to discretizations of surface integrals. In one dimension, our
simulations show that a model with the Wilson derivative and the Stratonovitch
prescription for this discretization leads to far better results at finite
lattice spacing than other models with Wilson fermions considered in the
literature. In particular, we check that fermionic and bosonic masses coincide
and the unbroken Ward identities are fulfilled to high accuracy. Equally good
results for the effective masses can be obtained in a model with the SLAC
derivative (even without improvement terms).\\
In two dimensions we introduce a non-standard Wilson term in such a way
that the discretization errors of the kinetic terms are only of order $O(a^2)$.
Masses extracted from the corresponding manifestly supersymmetric model prove
to approach their continuum values much quicker than those from a model
containing the standard Wilson term. Again, a comparable enhancement can
be achieved in a theory using the SLAC derivative.
\end{abstract}

{\small
}

\end{titlepage}
\setlength{\parindent}{0cm}
%\maketitle
\newlength{\origbaseskip}
\setlength{\origbaseskip}{\baselineskip}
\setlength{\baselineskip}{0.9\baselineskip}
\tableofcontents
\setlength{\baselineskip}{\origbaseskip}

\section{Introduction}
\label{sec:one}
Supersymmetry is an important ingredient of modern high energy physics beyond the standard model; 
since boson masses are protected by supersymmetry in such theories with chiral fermions, it helps
to reduce the hierarchy and fine-tuning problems drastically, and within grand unified theories, it leads to predictions of the proton life-time in agreement with present day experimental bounds. As low energy physics is manifestly not supersymmetric, this symmetry has to be broken at some energy scale. However, non-renormalization theorems in four dimensions ensure that tree level supersymmetric theories preserve supersymmetry at any finite order of perturbation theory; therefore, supersymmetry has to be broken non-perturbatively~\cite{Witten}.

The lattice formulation of quantum field theories provides a systematic tool to investigate non-perturbative problems. In the case of supersymmetric field theories, their formulation is hampered by the fact that the supersymmetry algebra closes on the generator of infinitesimal translations~\cite{NicolaiDondi}. Since Poincar{\'e} symmetry is explicitly broken by the discretization, one is tempted to modify the supersymmetry algebra so as to close on discrete translations. However, as lattice derivatives do not satisfy the Leibniz rule, supersymmetric actions for interacting theories will in general not be invariant under such lattice supersymmetries. The violation of the Leibniz rule is an $O(a)$ effect, and supersymmetry naively will therefore be restored in the continuum limit. In the case of Poincar{\'e} symmetry, the discrete remnants of the symmetry on the lattice are sufficient to prohibit the appearance of relevant operators in the effective action which are invariant only under a subset of the Poincar{\'e} group and require fine tuning of their coefficients in order to arrive at an invariant continuum limit. In generic lattice formulations, there are no discrete remnants of supersymmetry transformations on the lattice; in such theories, supersymmetry in the continuum limit can only be achieved by appropriately fine-tuning the bare couplings of all supersymmetry-breaking counterterms~\cite{BartelsKramer}.

An additional complication in the formulation of supersymmetric theories on the lattice is the fermion doubling problem. Local and translationally invariant hermitean Dirac operators on the lattice automatically describe fermions of both chiralities~\cite{NielsenNinomiya1, NielsenNinomiya2}; the fermionic extra degrees of freedom are usually not paired with bosonic modes and so lead to supersymmetry breaking. Generic prescriptions eliminating these extra fermionic modes also break supersymmetry.

As a simple supersymmetric theory, the Wess-Zumino model in two dimensions has been the subject of intensive analytic and numerical investigations. Early attempts included the choice of the nonlocal SLAC derivative (thereby avoiding the doubling problem) in the ${\cal N}=1$ version of this model~\cite{BartelsBronzan, Nojiri1, Nojiri2}. This was motivated by the idea that the Fourier transform of the lattice theory should coincide with that of the continuum theory. A Hamiltonian
approach where the SLAC derivative minimizes supersymmetry-breaking artifacts introduced by non-antisymmetric lattice derivatives was discussed in~\cite{KirchbergLaengeWipf}. Alternatively,
a local action with Wilson fermions was constructed at the cost of a nonlocal supersymmetry variation~\cite{GoltermanPetcher};\footnote{A similar approach with staggered fermions leads to problems in the continuum limit~\cite{BanksWindey}. -- In the four-dimensional model with ${\cal N}=1$ supersymmetry with Ginsparg-Wilson fermions, lattice chiral symmetry is incompatible with Yukawa couplings~\cite{FujikawaIshibashi1, FujikawaIshibashi2}; however, the theory can be regularized by supersymmetric higher derivative corrections, which leads to a supersymmetric continuum limit within perturbation theory~\cite{Fujikawa1}. It could be shown that supersymmetric Ward identities are satisfied up to order $O(g^2)$ in the coupling constant~\cite{BoniniFeo1, BoniniFeo2}.} simulations of this model~\cite{CatterallKaramov1} indicate that this theory with a cubic superpotential indeed features non-perturbative supersymmetry breaking.

In order to manifestly preserve some subalgebra of the ${\cal N}=1$ supersymmetry algebra on the lattice, the above discussion suggests to choose a subalgebra independent of the momentum operator.
Thus, supercharges $Q_+$ and $Q_-$ can be defined using (non)local derivatives on a spatial lattice. The choice of a continuous time then allows for a Hamiltonian defined by $H:=Q_+^2=Q_-^2$ which commutes with the supercharges~\cite{ElitzurRabSchwimmer, RittenbergYanki} and automatically contains a Wilson term. This strategy can be generalized to the ${\cal N}=2$ model on a spatial lattice; an analysis shows that a subalgebra admitting an $O(2)$ R-symmetry can be preserved~\cite{SakaiSakamoto}. Simulations for the ${\cal N}=1$ (see~\cite{RanftSchiller1}) and the ${\cal N}=2$ model~\cite{RanftSchiller2} have been done using the local Hamiltonian Monte-Carlo methods. Further simulations based on the Green function Monte-Carlo method for ${\cal N}=1$ indicate that supersymmetry is unbroken for a quartic superpotential and has a broken and an unbroken phase for cubic superpotentials~\cite{BeccariaRampino, BeccariaCampostrini}.

With a discrete time-coordinate, one has to resort to action- rather than Hamiltonian-based approaches. The perfect action approach was pursued for the free ${\cal N}=1$ theory in~\cite{Bietenholz}; a generalization for interacting theories seems however problematic. A treatment of the ${\cal N}=2$ Wess-Zumino model in the Dirac-K{\"a}hler formalism preserves a scalar supersymmetry on the lattice but leads to non-conjugate transformations of the complex scalar field and its conjugate. This enlarges the space of states and presumably renders the theory non-unitary~\cite{GoltermanPetcher}. A related~\cite{AratynBessaZimerman} idea avoiding these problems goes back to the idea of Nicolai~\cite{Nicolai} that (in this case) scalar supersymmetric field theories admit new bosonic variables with a Jacobian cancelling the determinant from integrating out the fermions, in terms of which the bosonic action is purely Gaussian. In general, the new variables are complicated nonlocal functions of the original ones; for ${\cal N}=2$ supersymmetry, however, local Nicolai variables can be expected~\cite{CecottiGirardello2}. Starting from a discretized form of the Nicolai variables found in the continuum, one simply defines the bosonic part of the action as the sum over squares of the Nicolai variables; the fermionic part of the action is adjusted so that the determinant of the fermion matrix still cancels the Jacobian. This leads to an action manifestly preserving part of the continuum supersymmetry on the lattice, and which contains improvement terms in addition to the naive latticization of the continuum action~\cite{CecottiGirardello2, SakaiSakamoto}. As stated in~\cite{ElitzurSchwimmer, CatterallKaramov2}, the breaking of some of the continuum supersymmetries can in this framework be traced back to the fact that the improvement terms are not compatible with reflection positivity. However, the violation of Osterwalder-Schrader positivity is an $O(a^2 g_{\rm phys})$ effect and thus should be negligible at least at weak coupling~\cite{Giedt}. An analysis of the perturbation series shows that these terms in an off-shell formulation of the theory with a cubic superpotential and Wilson fermions lead to tadpole diagrams which diverge linearly in the continuum limit~\cite{Fujikawa2}. A cancellation between these would-be central charge terms and the naive discretization of the continuum Hamiltonian has been suggested as a solution for the conundrum raised in~\cite{ElitzurSchwimmer} that the lattice result of the number of zero-modes of the Dirac operator seems to differ vastly from the continuum answer~\cite{KirchbergLaengeWipf}.

The possibility to introduce Nicolai variables is closely related to the fact that the $(2,2)$ Wess-Zumino model is a topological theory of Witten type, i.\,e., the action is of the form $S=Q\Lambda$ for a scalar supercharge $Q$, once auxiliary fields are introduced~\cite{BirminghamBlau, Catterall}. This formulation manifestly preserves $Q$-supersymmetry on the lattice and therefore guarantees that the theory remains supersymmetric in the limit of vanishing lattice spacing. Lattice theories of this type in various dimensions and with different degrees of supersymmetry have been classified in~\cite{GiedtPoppitz}. It turns out that in this formulation some remnant of the $U(1)_V$ R-symmetry which is left unbroken by the superpotential at the classical level but broken by the Wilson terms is restored in the continuum limit even non-perturbatively~\cite{Giedt}; this also indicates that at least for the cubic superpotential under consideration supersymmetry is not broken nonperturbatively.

At a fixed point in space, the Wess-Zumino model reduces to supersymmetric quantum mechanics. A naive discretization of the action with a Wilson term for just the fermion leads to differing masses for fermions and bosons in the continuum limit~\cite{CatterallGregory}. This can be traced back~\cite{GiedtKoniuk} to 1-loop contributions (of UV degree 0) of the fermion doublers to the boson propagator which have to be cancelled by an appropriate counterterm that was neglected in~\cite{CatterallGregory}. As soon as the action is amended by a Wilson term also for the boson~\cite{CatterallGregory}, an additional diagram involving the boson doublers cancels the finite correction of the fermion doublers, and no counterterms are required to achieve the desired continuum limit. In this case, fermion and boson masses agree in the limit of vanishing lattice spacing, and supersymmetric Ward identities are fulfilled to great accuracy.

As expected for a theory with ${\cal N}=2$ supersymmetry, supersymmetric quantum mechanics can be formulated in terms of local Nicolai variables~\cite{CecottiGirardello2}. However, this action differs from the amended form of~\cite{CatterallGregory} by a further interaction term which becomes an integral over a total derivative in the continuum limit.

In this paper, we scrutinize six different lattice actions for supersymmetric quantum
mechanics based on Wilson and SLAC fermions with and without such improvement terms.
The three models without improvement terms are not manifestly supersymmetric in the
presence of interactions and differ by bosonic terms which become irrelevant in the
continuum limit as well as by the choice of the lattice derivative; the other three
models with manifest supersymmetry differ in the prescription for the evaluation of
the improvement term and also in the choice of the lattice derivative --- they can
be constructed from three different Nicolai variables. We compare the effective
masses for interacting theories with a quartic superpotential and analyze the Ward
identities of the broken and unbroken supersymmetries at various lattice sizes and
couplings. The central observation here is that the manifestly supersymmetric theory
with the so-called Stratonovich prescription~\cite{EzawaKlauder} for the evaluation
of the improvement term (which is the discretization of a continuum surface integral)
leads to far better results than the model with an Ito prescription at finite lattice
spacing. The mass extraction for the models with SLAC derivative is at first sight
hampered by an oscillating behavior for nearby insertions in the bosonic and fermionic
two-point functions; however, this Gibbs phenomenon is under good analytical control
and can be softened by the application of an appropriate filter. With the help of
this ``optimal filter'', the results even surpass those of the model with Stratonovich
prescription.

Along the way, we show for which superpotentials and derivatives one can
guarantee positivity of the fermion determinants. In those cases where the determinant
can be computed exactly, we analyze under which circumstances they converge to
the correct continuum results; these exact results are crucial for our simulations.
Again, the model with Stratonovich prescription is ahead of the one with Ito prescription;
the former leads to a determinant with the correct continuum limit, whereas the latter
differs from it by a factor which depends on the superpotential.

Furthermore, we study three different manifestly supersymmetric discretizations of
the ${\cal N}{=}2$ Wess-Zumino model in two dimensions. Instead of trying to
generalize the Stratonovich prescription to two dimensions, we introduce a non-standard
Wilson term (corresponding to an imaginary Wilson parameter in the holomorphic
superpotential) in such a way that the discretization errors for the eigenvalues of
the free (bosonic and fermionic) kinetic operators are only of order $O(a^2)$ instead
of order $O(a)$ for the standard Wilson term. In the simulations, we study the effect
of the resulting violation of reflection positivity and compare the results with those
of the model with SLAC fermions. Due to calculational constraints, we have to restrict
the computations to smaller and intermediate values of the coupling.

As a theoretical background, we show that the discretized Wess-Zumino model in two
dimensions with the SLAC derivative has a renormalizable continuum limit.

The paper is organized as follows: In section~\ref{sec:two}, we introduce the quantum
mechanical models on the lattice, with and without improvement terms, and discuss which
behavior of the interacting theories at finite lattice spacings can be gathered from
their (non-)invariance under supersymmetry transformations in the free case. In
subsection~\ref{sec:deter}, we give details about the positivity of the fermion
determinants and derive their respective continuum limits. The results of the
effective masses and the Ward identities can be found in section~\ref{sec:simulqm}.
In section~\ref{sec:WZtwodim}, we discuss the discretizations of the ${\cal N}=2$
Wess-Zumino model in two dimensions and present the results of the mass extractions.
Section~\ref{sec:four} covers the algorithmic aspects of our simulations
including a derivation of the Gibbs phenomenon for the SLAC correlators which
justifies the application of the filter for the mass extraction in the quantum
mechanical model. Finally, section~\ref{sec:three} contains the proof that the
Wess-Zumino model on the lattice with the SLAC derivative is renormalizable to
first order in perturbation theory; a technical part of this proof is completed in appendix~\ref{sec:AppA}.
%%%%%%%%%%%%%%%%%%%%%%%%%%%%%%%%%%%%%%%%%%%%%%%

\section{Supersymmetric quantum mechanics}
\label{sec:two}
In this section we introduce the action for supersymmetric quantum mechanics
in a language which can be easily generalized later on to the two-dimensional
Euclidean Wess-Zumino model. In subsections
\ref{sec:unimproved} and \ref{sec:detail2} we present
six different lattice versions of the continuum theory with Euclidean action
\eqnl{
  S_{\rm cont} = \int d\tau \Big(\frac{1}{2}\dot{\phi}^2 + \frac{1}{2}W'^2
    + \psb\dot{\psi} + \psb W''\psi\Big),\quad\hbox{where}\quad
W'(\phi)\equiv\frac{dW(\phi)}{d\phi}.}{sqm1}
The continuum model is invariant under two supersymmetry transformations,
\begin{equation}
  \begin{array}{c@{\qquad}l@{\qquad}l}
    \delta^{(1)}\phi = \bar{\ve}\psi,& \delta^{(1)} \psb = -\bar{\ve}(\dot{\phi}+W'), & \delta^{(1)}\psi =0,\\
    \delta^{(2)}\phi = \psb\ve, & \delta^{(2)}\psb = 0, & \delta^{(2)}\psi = (\dot{\phi}-W')\ve
  \end{array}
  \label{sqm3}
\end{equation}
with anticommuting parameters $\ve$ and $\bar{\ve}$.
The lattice approximations considered below differ by the choice 
of the lattice derivative and/or the discretization 
prescription for a continuum surface term; in subsection~\ref{sec:deter} we 
argue why three of them lead to far better approximations to the continuum theory.
The results of our simulations for a quartic superpotential $W(\phi)=\frac{m}{2}\phi^2
+ \frac{g}{4}\phi^4$ (with positive $m$, $g$) are discussed in section~\ref{sec:simulqm}.

\subsection{Lattice models}\label{sec:QM}
We start from a one-dimensional periodic time lattice $\Lambda$
with real bosonic variables $\phi_x$ and two sets of real Gra{\ss}mann variables
$\psi_x$, $\psb_x$ on the lattice sites $x\in\Lambda=\{1,\dots,N\}$.
The integral and continuum derivative in \refs{sqm1} 
are replaced by a Riemann sum $a\sum$ and a lattice derivative $\pa$,
where $a$ denotes the lattice constant.
Two different \emph{antisymmetric} lattice derivatives will be 
used in what follows. These are the ultralocal derivative \eqnl{
\pah_{xy}=\frac{1}{2a}(\delta_{x+1,y}-\delta_{x-1,y})}{latder1}
with doublers\footnote{It should be noted that the \textit{symmetric}
combination of forward and backward derivatives leading to \refs{latder1}
yields an \textit{antisymmetric} matrix $(\pah_{xy})$.} and the nonlocal
SLAC derivative without doublers, which for an odd number $N$ of lattice
sites takes the form~\cite{Drell:1976bq, KirchbergLaengeWipf}
\eqnl{
\paslac_{x\neq y} =\frac{(-1)^{x-y}}{a}\frac{\pi/N}{\sin(\pi(x-y)/N)}\mtxt{and} \paslac_{xx}=0.}{latder3}
If we allow for non-antisymmetric derivatives then we may add
a multiple of the symmetric lattice Laplacian
\eqnl{
\Delta_{xy}=\frac{1}{a^2}\big(\delta_{x+1,y}-2\delta_{xy}+\delta_{x-1,y}\big)}
{latder5}
to $\pah$ to get rid of the doublers. In this way, we obtain a one-parameter
family of ultralocal derivatives with Wilson term,
\eqnl{
\pah-\ft12 ar\Delta,\quad -1\leq r\leq 1,}{latder7}
interpolating between the forward (or right) derivative 
for $r=-1$ and the backward (or left) derivative for $r=1$. As will be argued in
subsection~\ref{sec:deter}, for the quartic superpotential
$W(\phi)=\frac{m}{2}\phi^2 + \frac{g}{4}\phi^4$ with positive parameters $m$ and
$g$ to be considered below we shall need the latter with matrix elements
\eqnl{
\pab_{xy}=\pah_{xy}-\frac{a}{2}\Delta_{xy}=
\frac{1}{a}(\delta_{x,y}-\delta_{x-1,y}).}{latder9}
%%%%%%%%%%%%%%%%%%%%%%%%%%%
The backward derivative is free of doublers and not antisymmetric. For periodic 
boundary conditions the derivative operators $\paslac,\pah$ 
and $\pab$ are all given by circulant matrices
which commute with each other.

%%%%%%%%%%%%%%%%%%%%%%%%%%%%%%%
\subsection{Lattice models without improvement}
A straightforward discretization of the continuum action~\refs{sqm1} would be
\eqnl{
  S_{\rm naive} = \frac{a}{2} \sum_{x}\left((\pa\phi)_x^2 +
W_x^2\right) + a\sum_{x,y}\psb_x\,\big(\pa_{xy}
+W_{xy}\big)\,\psi_y,}{naiveaction}
where we scrutinize below three possibilities for the lattice derivatives $\pa$
as well as for the terms $W_x$, $W_{xy}$ derived from the superpotential\footnote{In
general, $W_x$ is not equal to $W'(\phi_x)$. The definition for each model can be
found below.} so that the theory is free of fermion doublers. None of these models
is supersymmetric under the discretization of any of the continuum supersymmetries~%
\refs{sqm3},
\begin{equation}
\begin{array}{c@{\qquad}l@{\qquad}l}
  \delta^{(1)}\phi_x = \bar{\ve}\psi_x,&
  \delta^{(1)}\psb_x = -\bar{\ve}\big((\pa\phi)_x+W_x\big),&
  \delta^{(1)}\psi_x = 0,\\
  \delta^{(2)}\phi_x =\psb_x\ve,&
  \delta^{(2)}\psb_x =0,&
  \delta^{(2)}\psi_x = \big((\pa{\phi})_x-W_x\big)\ve,
\end{array}
\label{lsqm13}
\end{equation}
however, at least for a free theory both supersymmetries are realized in two
of the models with antisymmetric matrices $(\pa_{xy})$. Thus, we might expect
a better approximation to the continuum theory for these models. In fact, it
will turn out in section~\ref{sec:simulqm} that this behavior pertains to the
interacting case, e.\,g., the masses extracted from these two models are
much closer to their continuum values than those from the third theory. Truly
supersymmetric (improved) lattice models will be considered in section~\ref{sec:improved}.
\subsubsection{The unimproved models in detail}\label{sec:unimproved}
\emph{(i) Naive lattice model with Wilson fermions}\\
The most naive discretization is given by the action~\refs{naiveaction}
with an additional Wilson term shifting the derivative $\pah$ as explained
in~\refs{latder9}, i.\,e.,
\eqnl{
S^{(1)}_{1d}=S_{\rm naive},\quad \pa=\pab,\quad W_x=W'(\phi_x)
\mtxt{and}W_{xy}=W''(\phi_x)\de_{xy}.}{model1}
The Wilson term removes fermionic as well as bosonic doublers. This action
has no supersymmetries at all, and bosonic and fermionic excitations have 
different masses in the continuum limit in the presence of interactions. Even
the free model has no exact supersymmetry; this can be traced back to the
fact that the derivative $\pab$ is not antisymmetric.\\[2mm]
%%%%%%%%%%%%%%%%%%%%
\emph{(ii) Naive lattice model with shifted superpotential}\\
An alternative way to remove the fermion doublers employed by Golterman and Petcher
and later Catterall and Gregory~\cite{GoltermanPetcher, CatterallGregory} is
to use the (unshifted) antisymmetric matrix $(\pah_{xy})$ and add a Wilson
term to the superpotential,
\eqnl{
S^{(2)}_{1d}=S_{\rm naive},\quad \pa=\pah,\quad W_x=-\frac{a}{2}(\Delta\phi)_x+W'(\phi_x),
\quad W_{xy}=-\frac{a}{2}\Delta_{xy}+W''(\phi_x)\de_{xy}.}{model2}
It should be noted that (as compared to \refs{model1}) only the bosonic
terms are changed. This model is only supersymmetric without interaction,
i.\,e, for $W'(\phi_x)=m\phi_x$. In the interacting case all susy Ward
identities are violated. The breaking is equally strong for both
supersymmetries.\\[2mm]
\emph{(iii) Naive lattice model with SLAC derivative}\\
Naively, one might expect the supersymmetry breaking effect in 
the naive lattice action \refs{naiveaction} with SLAC derivative
\refs{latder3} to be of the same magnitude as with backward derivative $\pab$.
Surprisingly enough, this is not so; it will turn out that the
mass extraction from the model
\eqnl{
S^{(3)}_{1d}=S_{\rm naive},\quad \pa=\paslac,\quad W_x=W'(\phi_x),\quad
W_{xy}=W''(\phi_x)\de_{xy}}{model3}
is about as good as for the improved actions considered below. This is again
related to the fact that the derivative is antisymmetric such that the free model
with SLAC derivative admits both supersymmetries (in contrast to the model 
with backward derivative).
%%%%%%%%%%%%%%%%%%%%%%%%%%%
\subsection{Lattice models with improvement}\label{sec:improved}
In order to preserve one of the two lattice supersymmetries in
\refs{lsqm13} for interacting theories the naive discretization
\refs{naiveaction} should be amended by extra terms which turn into
surface terms in the continuum limit and reinstall part of the continuum
supersymmetry on the lattice. Such invariant lattice models may be
constructed with the help of a Nicolai map $\phi\mapsto\xi(\phi)$ of the
bosonic variables. In terms of the Nicolai variables $\xi_x(\phi)$ the improved 
lattice actions take the simple form
\eqnl{
  S_{\rm susy} = \frac{a}{2}\sum_x \xi_x(\phi)^2
    + a\sum_{x,y}\psb_x\,\frac{\pa\xi_x}{\pa\phi_y}\,\psi_y .}{impact1}
This is a discretization of the most general supersymmetric action in 
terms of a real bosonic variable and two real Gra{\ss}mann variables on the
circle~\cite{Nicolai}. It is easily seen to be invariant under the first
type of transformations
\eqnl{
  \delta^{(1)}\phi_x = \bar{\ve}\psi_x,\quad \delta^{(1)}\psb_x = -\bar{\ve}\xi_x,\quad
    \delta^{(1)}\psi_x = 0 .}{impact3}
For the particular choice $\xi_x(\phi)=(\pa\phi)_x +W_x$ the supersymmetric 
action~\refs{impact1} becomes
\eqngrl{
  S_{\rm susy} &=& 
\frac{a}{2} \sum_x \left((\pa\phi)_x +W_x\right)^2 + a
\sum_{x,y}\psb_x\:\!\left(\pa_{xy} + W_{xy}\right)\psi_y}
{&=&
S_{\rm naive}+a\sum_x W_x(\phi)\,(\pa\phi)_x}{impact5}
and the supersymmetry transformation \refs{impact3} is identical
to $\de^{(1)}$ in~\refs{lsqm13}. So the improved model
\refs{impact5} is invariant under the first supersymmetry
$\delta^{(1)}$ for arbitrary superpotentials.
It differs from the naive discretization \refs{naiveaction} of the
continuum action \refs{sqm1} by the \emph{improvement term}
$a\sum_x W_x\,(\pa\phi)_x$ which turns into an integral over
a total derivative in the continuum, and hence zero for periodic
boundary conditions. The improvement term is needed for an
invariance of the lattice action under one supersymmetry
transformation. For an interacting theory the lattice action~%
\refs{impact5} is not invariant under the other supersymmetry
transformation $\delta^{(2)}$ with parameter $\ve$; an action
preserving only this symmetry can be analogously constructed,
\eqnl{
\tilde S_{\rm susy} = \frac{a}{2}\sum_x \tilde\xi_x(\phi)^2
    + a\sum_{x,y}\psb_x\,\frac{\pa\tilde\xi_y}{\pa\phi_x}\,\psi_y .}{impact7}
It is invariant under the nilpotent supersymmetry transformations
\eqnl{
  \delta^{(2)}\phi_x = \psb_x\ve,\quad \delta^{(2)}\psb_x = 0, \quad
    \delta^{(2)}\psi_x = -\tilde\xi_x \ve.}{impact9}
To generate the same fermionic term as in $S_{\rm susy}$ we choose
$\tilde\xi_x=-(\pa\phi)_x+\Wt_x$ with antisymmetric $(\pa_{xy})$ and symmetric 
$W_{xy}$. Then the supersymmetry \refs{impact9} agrees with $\delta^{(2)}$ 
in \refs{lsqm13}, and the action takes the form
\eqngrl{
\tilde S_{\rm susy}&=&
\frac{a}{2} \sum_x \big((\pa\phi)_x -W_x\big)^2 + 
a\sum_{x,y}\psb_x\,\big(\pa_{xy} +W_{xy}\big)\,\psi_y}
{&=& S_{\rm naive}-a\sum_x W_x(\phi)\,(\pa\phi)_x
}{impact11}
For periodic fields $S_{\rm susy}$ and $\tilde S_{\rm susy}$ converge to the
same continuum limit. On the lattice they are only equal in the noninteracting
case.\footnote{In order to preserve the second supersymmetry $\delta^{(2)}$
also for $S_{\rm susy}$ in~\refs{impact5} in the absence of interactions, its
definition will have to be slightly modified only for the Stratonovich
prescription to be discussed below.}
%%%%%%%%%%%%%%%%%%%%%%%%%%%%%
\subsubsection{The improved models in detail}\label{sec:detail2}
We consider three supersymmetric versions of the discretization~\refs{impact5}
with improvement term.\\[2mm]
\emph{(iv) Supersymmetric model with Wilson fermions and Ito prescription}\\
In order to avoid doublers and at the same time keep half of supersymmetry we use 
the antisymmetric matrix $(\pah_{xy})$ and shift the superpotential by a
Wilson term~\cite{BeccariaCurciDAmbrosio}.\footnote{It is obvious that this is
equivalent to working with a shifted lattice derivative as in~\refs{latder9}
and an unshifted superpotential since the action now only depends on the
invariant combination $\xi_x$.} The corresponding model
\eqnl{
S^{(4)}_{1d}=S_{\rm susy},\quad \pa=\pah,\quad W_x=-\frac{a}{2}(\Delta \phi)_x+
W'(\phi_x),\quad W_{xy}=-\frac{a}{2}\Delta_{xy}+W''(\phi_x)\de_{xy}}{model4}
is invariant under the supersymmetry $\delta^{(1)}$~\cite{CatterallGregory}.
Of course, the non-interacting model is also invariant under $\delta^{(2)}$.

With these definitions, the improvement term is given by the well-known Ito
prescription $\sum_x W'(\phi_x)\,(\phi_x-\phi_{x-1})$.\\[2mm]
\emph{(v) Supersymmetric models with Wilson fermions and Stratonovich prescription}\\
Instead of the Ito prescription, we can choose the Stratonovich scheme~%
\cite{BeccariaCurciDAmbrosio} for the evaluation of the surface term, $\sum_x W'(\sigma_x)\,(\phi_x-\phi_{x-1})$ with $\sigma_x=\half(\phi_x+\phi_{x-1})$.\footnote{For
monomial superpotentials $W(\phi)=\phi^k,\;k=1,2,\dots$, this prescription is equivalent
to the prescription $\sum_x \frac{1}{2}\big( W'(\phi_x)+W'(\phi_{x-1})\big)\,
(\phi_x-\phi_{x-1})$ ; in the latter case, the superpotential terms are evaluated
only at a given lattice site.} The corresponding action can be obtained from
\refs{impact1} with a Nicolai variable $\xi_x(\phi)=(\pab\phi)_x + W'(\sigma_x)$,
\eqnl{
S^{(5)}_{1d}=S_{\rm susy},\quad \pa=\pah,\quad W_x=-\frac{a}{2}(\Delta\phi)_x+W'(\sigma_x),\quad
W_{xy}=-\frac{a}{2}\Delta_{xy}+\frac{\pa W'(\sigma_x)}{\pa\phi_y}.
}{model5}
One should note that this procedure differs from the one proposed in~\cite{EzawaKlauder},
where the fermions are first integrated out in the continuum theory, and only
then a Stratonovich interpretation is given for the surface term -- in this case,
the fermionic path integral of the Euclidean evolution operator has to be defined
in a non-standard way in order for the bosonic Stratonovich Jacobian to cancel
the fermion determinant.

We will see that compared to the fermion determinant involving continuum
derivatives, the continuum limit of the fermion determinant for the Ito
prescription is off by a factor depending on the superpotential whereas the
Stratonovich prescription reproduces exactly the desired continuum result.

Since~\refs{model5} was constructed from~\refs{impact1}
in terms of Nicolai variables, it is manifestly supersymmetric under $\delta^{(1)}$
as given in~\refs{impact3}. The discretization of the second supersymmetry
is in general not preserved on the lattice, not even in the free case. This
latter fact suggests a modification of $\tilde\xi_x(\phi)$ in~\refs{impact9} to
\eqnl{
    \tilde\xi_x(\phi) = -(\pah\phi)_x-\frac{a}{2}(\Delta\phi)_x+W'(\sigma_x').}{model5a}
This changes effectively the backward- into a forward-derivative, and the
derivative of the superpotential is evaluated now at
$\sigma_x'=\frac{1}{2}(\phi_x+\phi_{x+1})$. With these definitions, $\delta^{(2)}$
is a symmetry of the action~\refs{model5} in the absence of interactions. It is
also this variation with which we compute Ward identities in section~\ref{sec:simulqm}.
\\[2mm]
\emph{(vi) Supersymmetric models with SLAC derivative}\\
In order to avoid fermion doublers, we can specialize $\pa$
to be the SLAC derivative,
\eqnl{
S^{(6)}_{1d}=S_{\rm susy},\quad \pa=\paslac,\quad W_x=W'(\phi_x).}{model6}
In spite of its nonlocality, the fermion and boson masses extracted from
two-point functions prove to approach the continuum value quite fast; the
quality turns out to be comparable to that of the Stratonovich prescription.
The interacting supersymmetric model with SLAC derivative is only invariant
under $\de^{(1)}$ in~\refs{impact3} by construction.
%%%%%%%%%%%%%%%%%
\subsection{Fermion determinants}
\label{sec:deter}
In this subsection we demonstrate which sign of the Wilson term
we must choose in order to guarantee positivity of
the fermion determinant. After that, it will become clear that as compared
to the value for the continuum operator, the fermion determinant for the Ito
prescription is off by a factor depending on the superpotential whereas the
Stratonovich prescription and the SLAC derivative reproduce the desired
continuum result.
\subsubsection{Sign of the determinants}
For a real fermion matrix $\pa_{xy}+W_{xy}$, complex eigenvalues $\l$
appear pairwise as $\l\bar{\l}$ in the determinant. Hence, the determinant can only
become negative through real eigenvalues. Since without loss of generality,
eigenvectors $v_x$ to real eigenvalues $\l$ can be taken to be real (otherwise,
take $v_x + v_x^*$) and normalized, only the symmetric part
of the fermion matrix contributes to a real $\l$:
\eqnl{
 \l = \sum_{x,y}v_x\left(\pa_{xy}+W_{xy}\right)v_y = \sum_{x,y}v_x
 \left(\pa^{\,\rm s}_{xy} + W_{xy}\right)v_y.}{detsign1}
For the antisymmetric \emph{SLAC derivative} the symmetric part $\pa^{\rm s}$
is absent and no Wilson term is required, $W_{xy}=W'(\phi_x)\delta_{xy}$,
such that the real eigenvalues are given by
\eqnl{
\l=\sum_x W''(\phi_x)v^2_x.}{detsign3}
We conclude that all real eigenvalues and thus the determinant will be 
positive for the models \emph{(iii)} and  \emph{(vi)} in case $W''$ is nonnegative
definite. For the models \emph{(i), (ii)} and \emph{(iv)} \emph{with Wilson term}
the real eigenvalues are given by
\eqnl{
\l=-\frac{ar}{2}\sum_{xy}v_x\Delta_{xy}v_y+\sum_v W''(\phi_x)v^2_x.}
{detsign5}
Since $-\Delta$ is positive all real eigenvalues $\lambda$ and therefore the
determinant will be positive definite in case the $W''(\phi_x)$ are nonnegative
and $r>0$; as usual, in this paper we choose $r=1$ in this case. Vice versa, for
a negative $W''(\phi_x)$ we would have to change the sign of the Wilson term
(i.\,e., choose $r=-1$) for the fermionic determinant to stay positive.\footnote{Either
way, this part of the action satisfies site- as well as link reflection positivity~\cite{MontvayMuenster}.}
In the following subsection we shall prove by an explicit calculation that also
for the model \emph{(v)} with Stratonovich prescription the fermionic determinant
is positive for positive $W''$.

If $W''$ is neither positive nor negative definite positivity of the fermion
determinant is not expected. In fact, for instance
in the particular example of $W$ (and $W''$) being an odd power of $\phi$ we
have to expect a change of sign: From the interpretation of the Witten index
\eqnl{
  Z_{\rm W} = \int_{\rm per. b.c.} D\psb D\psi D\phi\, e^{-S}
    = \int D\phi \det(\pa + W''(\phi))\, e^{-S_{\rm bos}}.}{detsign7}
(for the path integral with periodic boundary conditions for all fields)
as the winding number of the Nicolai map $\xi=\pa\phi + W'$ regarded as a map
from the space of bosonic variables to itself~\cite{CecottiGirardello1}, we
expect that there may be phases with broken supersymmetry for $W'$ even; the
Witten index vanishes. This would be impossible for a determinant with
definite sign.
\subsubsection{Calculating the determinants}
The fermionic determinant for the Ito and Stratonovich prescription can be
computed exactly. The (regularized) determinant of the \emph{continuum operator}
$\pa_\tau + W''(\phi(\tau))$ on a circle of radius $\beta$ can easily be seen to
be~\cite{Gozzi}
\eqnl{
  \det\left(\frac{\pa_\tau + W''(\phi(\tau))}{\pa_\tau + m}\right) = 
    \frac{\sinh\big(\frac{1}{2}\int_0^\beta d\tau\,
    W''(\phi(\tau))\big)}{\sinh(\frac{\beta}{2}m)};}{det1}
this is the value with which we have to compare the lattice results.
Note that for a non-negative $W''$ the determinant is positive.

For \emph{Ito's calculus} with Wilson fermions, the ratio of the 
determinant of the fermion matrix for the interacting theory
\eqnl{
\pa_{xy} + W_{xy}%_{\rm Ito} 
=\pab_{xy}+W''(\phi_x)\delta_{xy}}{det3}
to that of the free theory is given by
\eqnl{
  \det\left(\frac{\pa + (W_{xy})}{\pab + m\one}\right)_{\rm Ito}
    = \frac{\prod\big(1 + aW''(\phi_x)\big)-1}{(1+am)^N - 1}.}{det5}
It is positive for positive $W''(\phi)$ and this agrees with
the results in the previous section.
For $N=\beta/a\to\infty$, the product converges 
to\footnote{In a completely analogous manner, the right-derivative would
lead to the inverse prefactor in front of the continuum result.}
\eqnl{
  \det\left(\frac{\pa + (W_{xy})}{\pab + m\one}\right)_{\rm Ito}
\stackrel{N\to\infty}{\longrightarrow}\;
\frac{e^{\int_0^\beta d\tau\,W''(\phi(\tau))\,d\tau/2}}{e^{\beta m/2}}\,
  \det\left(\frac{\pa_\tau + W''(\phi(\tau))}{\pa_\tau + m}\right)}{det7}
since $\ln\prod_x(1+a W''(\phi_x))=\sum\ln(1+a W''(\phi_x))\to\int dx\,W''(\phi_x)$.
The limit~\refs{det7} differs by a field dependent factor from the continuum result.

For Wilson fermions with \emph{Stratonovich prescription} the 
regularized determinant of the fermion matrix
\eqnl{
 \pa_{xy} + W_{xy} =
\pab_{xy}+\half W''(\sigma_x)\left(\delta_{xy}+\delta_{x-1,y}\right)}{det9}
is again positive for positive $W''$. 
But in contrast to the determinant \refs{det5} with Ito prescription it
converges to the continuum result,
\eqnl{
  \det\left(\frac{\pa + (W_{xy})}{\pab + (m_{xy})}\right)_{\rm Strat}
    = \frac{\prod(1+\frac{a}{2}M_x)
    -\prod (1-\frac{a}{2}M_x)}{\prod(1+\frac{a}{2}m)-\prod(1-\frac{a}{2}m)}
    \stackrel{N\to\infty}{\longrightarrow}\det\left(\frac{\pa_\tau +
    W''(\phi(\tau))}{\pa_\tau +m}\right)\!\!,}{det11}
where $m_{xy}=\frac{1}{2}(\delta_{xy}+\delta_{x-1,y})$ and $M_x=W''(\sigma_x)$.
One can show that for $N\to \infty$ the fermionic determinant with SLAC derivative
converges rapidly to the continuum result.
%%%%%%%%%%%%%%%%%%%%%%%%%%%%%%%%%%%%%%%%%%%%%%

\section{Simulations of supersymmetric quantum mechanics}\label{sec:simulqm}
We have performed high precision Monte-Carlo simulations to investigate the
quality of the six lattice approximations introduced in 
subsections \ref{sec:unimproved} and  \ref{sec:detail2}. 
The models with actions $S_{1d}^{(1)},S_{1d}^{(2)}$ and $S_{1d}^{(3)}$
are not supersymmetric whereas the actions
$S_{1d}^{(4)},S_{1d}^{(5)}$ and $S_{1d}^{(6)}$
preserve the supersymmetry $\delta^{(1)}$.
\subsection{Effective Masses on the lattice}\label{sec:qmmasses}
In order to determine the masses we have calculated the fermionic and bosonic
two point functions
\eqnl{
G_{\rm bos}^{(n)}(x)=\langle \phi_x\phi_0\rangle\mtxt{and}
G_{\rm ferm}^{(n)}(x)=\langle \psb_x\psi_0\rangle,}{massen1}
in all models $(n)$ and fitted their logarithms with a
linear function. This way of determining the masses
$m_{\rm bos}(a)$ and $m_{\rm ferm}(a)$ from the slope of the
linear fit works well for all models with ultralocal derivatives.
Details on technical aspects of the extraction of effective masses can
be found in section~\ref{sec:filter}.
%%%%%%%%%%%%%%%%%%%%%%%%%%%%
\subsubsection{Models without interaction}\label{sec:massesqmfree}
The actions of the non-interacting models are quadratic
in the field variables,
\eqnl{
S_{\rm free}=\half \sum_{xy}\phi_x K_{xy}\phi_y+\sum_{xy}
\psb_x M_{xy}\psi_y.}{freem1}
Actually, for $W'(\phi)=m\phi$ all the improvement terms 
vanish and only four of the six actions introduced in the last
section are different. The corresponding matrices
$M$ and $K$ are given in the following list:\footnote{Note that
$\pab+(\pab)^T=-\Delta$ and $-\pah^2+\ft{1}{4}a^2
\Delta^2=-\Delta$.}
\eqnl{
\begin{array}{|c||c|c|c|c|}\hline
&S^{(1)}_{\rm free}& S^{(4)}_{\rm free}=S^{(2)}_{\rm free}& S^{(5)} _{\rm free}& S^{(6)}_{\rm free}=S^{(3)}_{\rm free}\\ \hline
M&\pab+m& \pab+m & (1-\frac{am}{2})\pab+m & \paslac+m\\
K&-\Delta+m^2 & -\Delta+m^2-am\Delta &
-\left(1-(\frac{am}{2})^2\right)\Delta+m^2& -(\paslac)^2+m^2\\
&&=M^TM&=M^TM&=M^TM\\ \hline
\end{array}
}{freem3}
For the actions in the last three columns we have $\det K=(\det M)^2$
as required by supersymmetry. This is not true for the non-supersymmetric
naive model $S^{(1)}_{\rm free}$ with Wilson fermions.

Without interactions, the masses of all models as determined by numerically
inverting $M$ and $K$ converge to the same continuum limit $m$. 
For the free supersymmetric models \emph{(ii)}-\emph{(vi)},
$m_{\rm bos}(a)$ and $m_{\rm ferm}(a)$ roughly
coincide even for finite lattice spacings. This is
to be expected for supersymmetric theories. In the first model,
$m_{\rm bos}(a)$ for finite $a$ is already very close to its
continuum value, in contrast to $m_{\rm ferm}(a)$ which for finite $a$ is
considerably smaller than $m_{\rm bos}(a)$.
Since the free supersymmetric action $S^{(4)}_{\rm free}$ has
the same fermionic mass as $S^{(1)}_{\rm free}$ and since
$m_{\rm bos}(a)\approx m_{\rm ferm}(a)$ for this
model, we conclude that its bosonic mass for finite $a$ is notably
smaller than its value in the continuum limit. Thus, when
supersymmetrizing the naive model with Wilson fermions we pay a
price: the boson masses get worse while approaching the fermion masses.

\begin{figure}[!ht]
\begin{center}
  \includegraphics{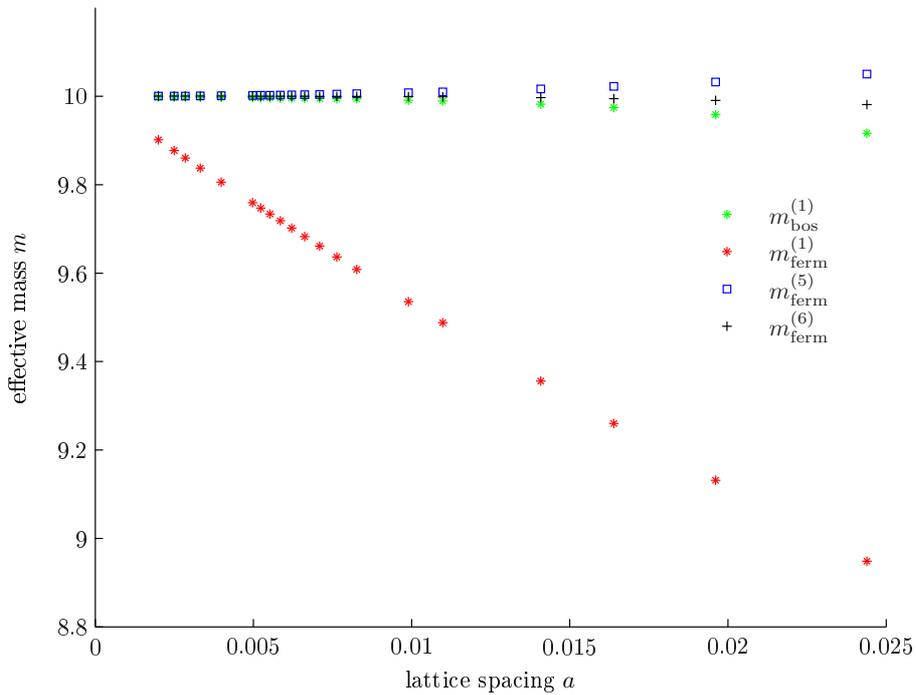}
  \caption{The masses determined by numerically inverting the kinetic
operators for the free theories. There are only $4$ different masses,
cf.~\refs{freem3}.}
  \label{fig_free:1}
\end{center}
\end{figure}

The situation is much better for the other
supersymmetric models with SLAC derivative or Wilson fermions
with Stratonovich prescription. The masses are equal and very close
to the continuum result already for finite $a$.
The masses for the Stratonovich prescription are comparable
to the boson masses of the naive model without supersymmetry.
The masses for the free models and their dependence on $a$
are depicted in figure~\ref{fig_free:1}.

\begin{figure}[!ht]
\begin{center}
  \resizebox{10.5cm}{!}{\includegraphics{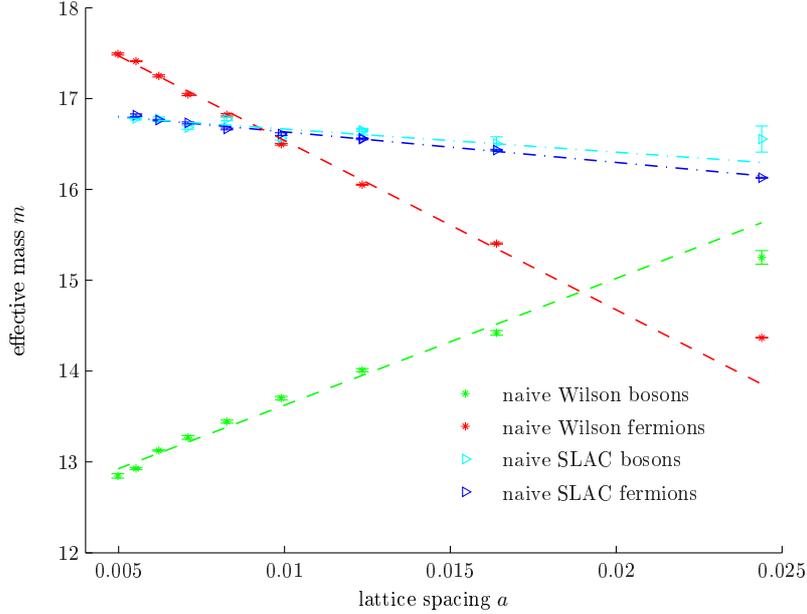}}
  \caption{Boson and fermion masses for Wilson fermions without improvement
(model \emph{(i)}) and non-supersymmetric SLAC fermions (model \emph{(iii)}).
The parameters for the linear fits can be found in table~\ref{intm7}.}
  \label{fig:1}
\end{center}
\end{figure}
From \refs{freem3} we conclude that
\eqnl{
G_{\rm ferm}^{(4)}(x,m)=\left(1-\frac{am}{2}\right)G^{(5)}_{\rm ferm}
\left(x,m\left(1-\frac{am}{2}\right)\right).}{freem5} 
The mass $m^{(5)}_{\rm ferm}(a)$ extracted from $G^{(5)}_{\rm ferm}$
is close to the continuum value $m$ such that
\eqnl{
m^{(1,2,4)}_{\rm ferm}(a)=m^{(2,4)}_{\rm bos}(a)
\approx m\left(1-\frac{am}{2}\right),}{freem6}
and this simple relation explains why the linear fit through the
masses $m^{(1)}_{\rm ferm}(a)$ marked with red dots in figure~\ref{fig_free:1}
has such a large negative slope.
%%%%%%%%%%%%%%%%%%%%%%%%%%%%%
\subsubsection{Models with interaction}
We have calculated the masses for the interacting models with
even superpotential
\eqnl{
W(\phi)=\frac{m}{2}\phi^2+\frac{g}{4}\phi^4\quad\Longrightarrow\quad
W'(\phi)^2=m^2\phi^2+2mg\phi^4+g^2\phi^6.}{intm1}
Since in the weak coupling regime the results are comparable
to those of the free models we have simulated the models at strong
coupling. In order to compare our results with those of
Catterall and Gregory in~\cite{CatterallGregory} we have picked
their values $m=10$ and $g=100$ for which the dimensionless 
ratio $g/m^2$ equals unity.
The energy of the lowest excited state has been calculated by
diagonalizing the Hamiltonian on large lattices with small $a$ and alternatively
with the shooting method. With both methods we obtain the continuum 
value
\eqnl{
m_{\rm phys}=16.865.}{intm3}
We now summarize the results of our MC simulations.
As in the free case, $m_{\rm bos}(a)\neq m_{\rm ferm}(a)$ for the
non-supersymmetric model with action $S^{(1)}_{\rm 1d}$, see figure \ref{fig:1}.
\begin{figure}[!p]
\begin{center}
  \resizebox{10.5cm}{!}{\includegraphics{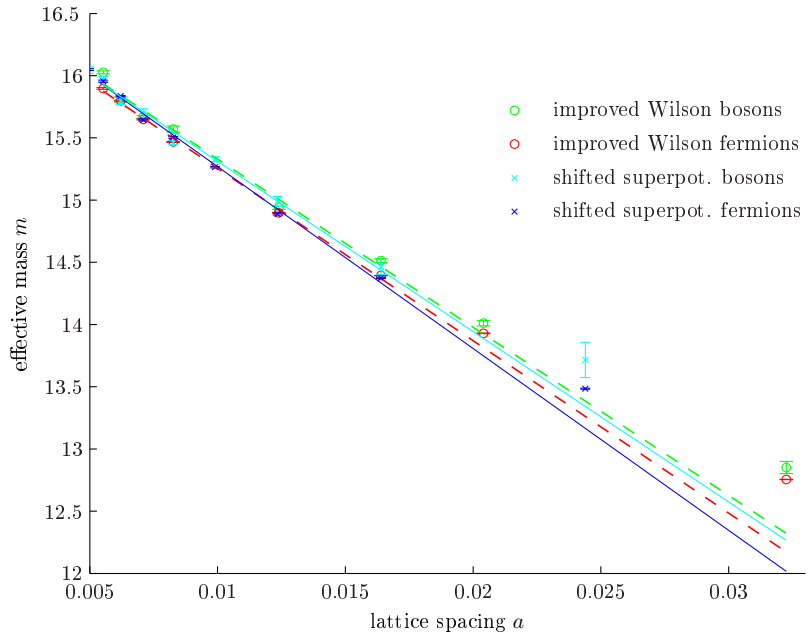}}
  \caption{The masses $m_{\rm bos}(a)$ and $m_{\rm ferm}(a)$ for
models \emph{(ii)} (shifted superpotential) and \emph{(iv)} (Ito improvement).}
  \label{fig:2}
\end{center}
\end{figure}
\begin{figure}[!p]
\begin{center}
  \resizebox{10.5cm}{!}{\includegraphics{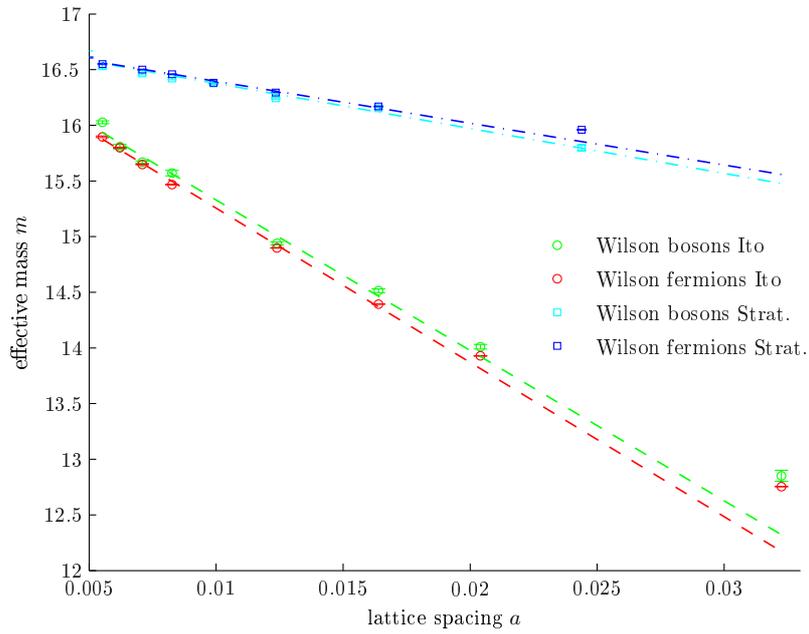}}
  \caption{Masses for supersymmetric models \emph{(vi)} (Ito prescription)
and \emph{(v)} (Stratonovich prescription). Only lattices with at least
61~sites are included in the fit.}
  \label{fig:3}
\end{center}
\end{figure}
In addition, for $g\neq 0$ their continuum values are different and none
of the two values agrees with \refs{intm3}. This has been
predicted earlier by Giedt et al.~\cite{GiedtKoniuk}. We conclude that the
naive lattice model with Wilson fermions is not supersymmetric for
$a\to 0$.

The Monte-Carlo results are much better for the second model 
with action $S^{(2)}_{\rm 1d}$ as given in \refs{model2}.
Although this model is not supersymmetric, its boson and fermion
masses are almost equal for finite lattice spacings. Linear extrapolations
to vanishing $a$ yield $m^{(2)}_{\rm bos}(0)=16.68\pm 0.05$ and 
$m^{(2)}_{\rm ferm}(0)=16.73\pm 0.04$ which are quite close to the correct value 
$16.865$. The results for the non-supersymmetric model with action
$S^{(2)}_{1d}$ and the supersymmetric model with action $S^{(4)}_{1d}$
are almost identical, similarly as for the free models. The masses for various
lattice constants between $0.005$ and $0.03$ for the two models are depicted
in figure \ref{fig:2}.
\begin{figure}[!ht]
\begin{center}
  \resizebox{10.5cm}{!}{\includegraphics{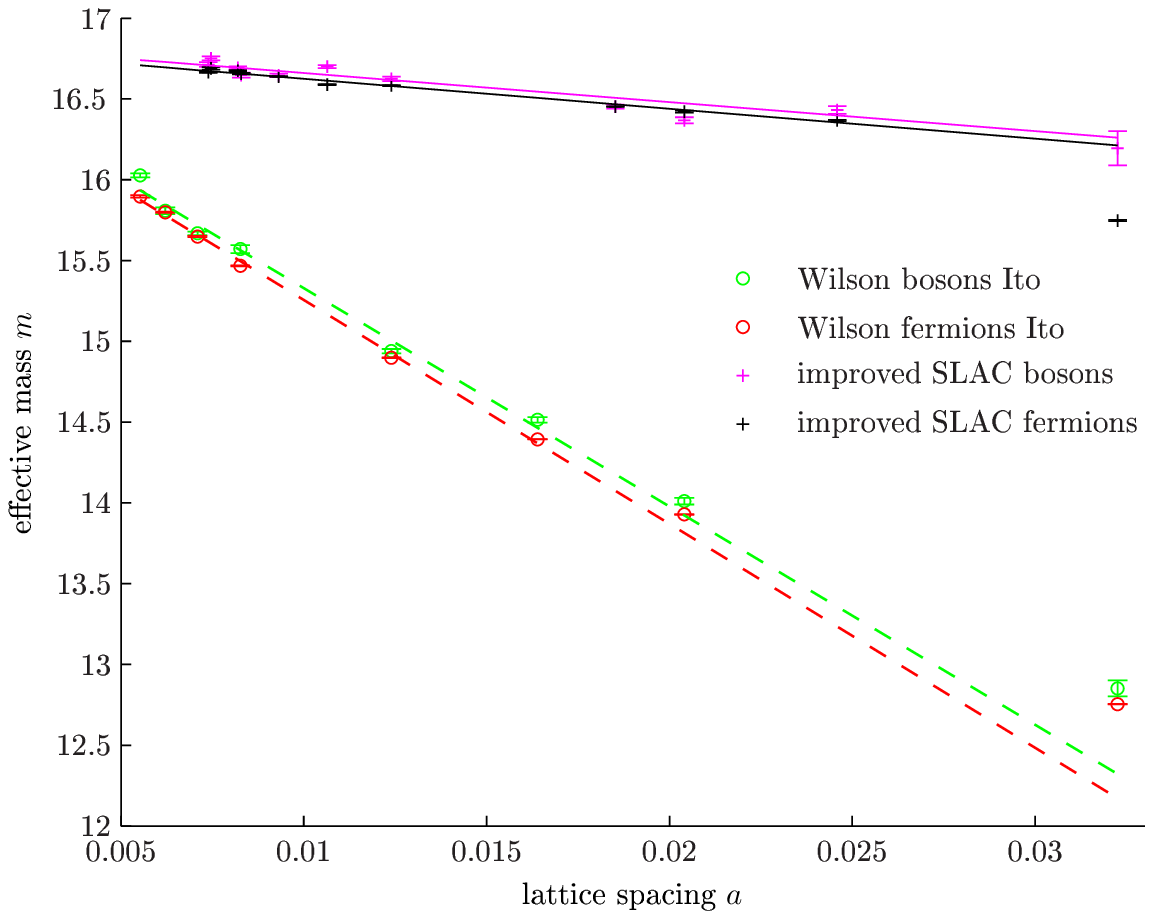}}
  \caption{Masses for the supersymmetric model \emph{(vi)} with SLAC derivative as
compared to the masses of the model \emph{(vi)} with Wilson fermions and Ito
prescription.}
  \label{fig:4}
\end{center}
\end{figure}
The corrections to the continuum value are of order $O(a)$ and are
as big as for the corresponding free models. The slope and intercepts
for the linear fits are listed in table \ref{intm7}.

At finite lattice spacing $a$, the masses $m_{\rm bos,ferm}(a)$
for model \emph{(v)} with Wilson fermions and Stratonovich prescription
are much closer to their respective continuum limits than for the model
\emph{(iv)} with Ito prescription. Furthermore, the extrapolated continuum
masses, $m^{(5)}_{\rm bos}(0)=16.78\pm 0.04$ and $m^{(5)}_{\rm ferm}(0)=
16.77\pm 0.02$, are very close to the correct value \refs{intm3}. The data
points for the supersymmetric models with Wilson fermions are depicted in
figure~\ref{fig:3}. Again the slope and intercepts of the linear fits can
be found in table \ref{intm7}. Of all lattice models with ultralocal
derivatives considered in this paper this model yields the best
predictions.

The model with Stratonovich prescription for the improvement term is 
outperformed only by the models \emph{(iii)} and \emph{(vi)} with nonlocal
SLAC derivative for fermions and bosons. This observation is not surprising,
since the remarkably high numerical precision of supersymmetric lattice models
with SLAC derivative has been demonstrated earlier in the Hamiltonian approach
in \cite{KirchbergLaengeWipf}. Furthermore, this is in line with our results
for the free models, see figure~\ref{fig_free:1}. The masses for the interacting
unimproved model~\emph{(iii)} are plotted in figure~\ref{fig:1} and those for
the improved model~\emph{(vi)} in figure \ref{fig:4}. Even for moderate lattice
spacings the masses $m_{\rm bos}(a)\approx m_{\rm ferm}(a)$ are very close to
their continuum limits $m_{\rm bos}^{(6)}(0)=16.84\pm 0.03$ and
$m_{\rm ferm}^{(6)}(0)=16.81\pm 0.01$ which in turn are off the true value
$16.865$ by only some tenth of a percent. The extrapolated masses for the
unimproved lattice model in table \ref{intm7} have a comparable precision.
But of course there is no free lunch, since for the SLAC derivative one must
smooth the two-point functions $G_{\rm bos,ferm}(x)$ with a suitable 
filter for a sensible mass extraction. Details on the filtering can be found in 
subsection \ref{sec:filter}.

In the following table we list the slopes $k$ and intercepts $m(0)$ of the linear fits
\eqnl{
m_{\rm bos}(a)=k_{\rm bos}\cdot a+m_{\rm bos}(0)\mtxt{and}
m_{\rm ferm}(a)=k_{\rm ferm}\cdot a+m_{\rm ferm}(0)} {intm5}
to the measured masses for the six lattice models considered.

\begin{figure}[!ht]
\begin{center}
  \resizebox{10.5cm}{!}{\includegraphics{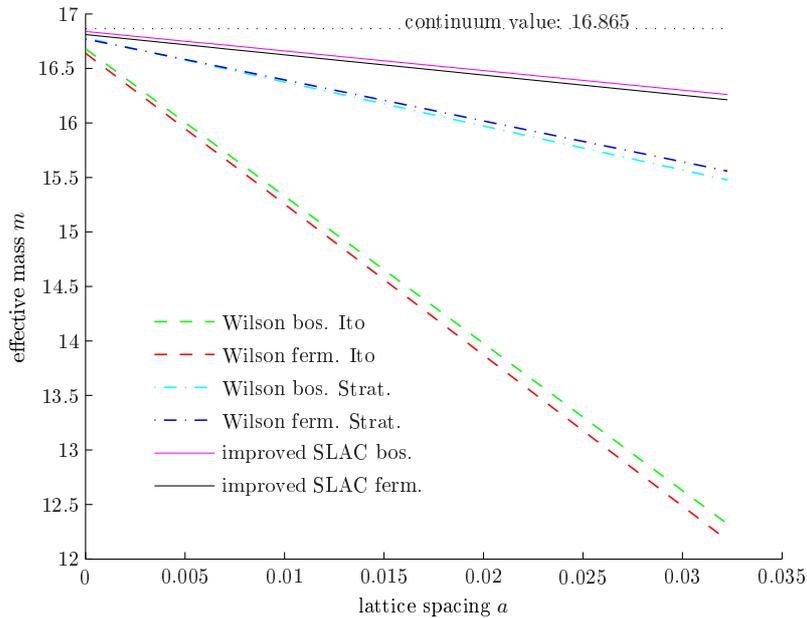}}
  \caption{Linear fits to the masses of all three supersymmetric lattice models.}
  \label{fig:5}
\end{center}
\end{figure}

\begin{table}
\begin{center}
\begin{tabular}{|c|rr|rr|}\hline
model &$ k_{\rm bos}\qquad$ & $m_{\rm bos}(0) \quad$ & $ k_{\rm ferm}\qquad
$ & $m_{\rm ferm}(0)\quad  $\\ \hline
$S^{(1)}_{1d}$ & $139.52\pm 8.45$ & $12.23\pm 0.08$ & $-186.25\pm 4.98$ & $18.40\pm 0.05 $\\
$S^{(2)}_{1d}$ & $-136.85\pm 5.22$ & $16.68\pm 0.05$ & $-146.10\pm 3.84$ & $16.73\pm 0.04$\\
$S^{(3)}_{1d}$ & $-25.22\pm 6.24$ & $16.92\pm 0.07$ & $-33.64\pm 2.52$ & $16.97\pm 0.03$\\ \hline
$S^{(4)}_{1d}$ & $-135.11\pm 7.36$ & $16.68\pm 0.07$ & $-138.50\pm 2.85$ & $16.64\pm 0.03$\\
$S^{(5)}_{1d}$ & $-40.40\pm 4.46$ & $16.78\pm 0.04$ & $-37.55\pm 1.98$ & $16.77\pm 0.02$\\
$S^{(6)}_{1d}$ & $-17.97\pm 2.41$ & $16.84\pm 0.03$ & $-18.53\pm 0.91$ & $16.81\pm 0.01$\\ \hline
\end{tabular}
\caption{Slope and intercepts of linear interpolations
for the masses.}\label{intm7}
\end{center}
\end{table}
The linear fits for the models with improvement \emph{(iii)}-\emph{(vi)}
are compared in figure~\ref{fig:5}. Lattice supersymmetry guarantees that
the boson and fermion masses are equal for these models.

\subsection{Ward identities}\label{sec:ward}
The invariance of the path integral measure under supersymmetry transformations
leads to a set of Ward identities connecting bosonic and fermionic correlation
functions. Namely, the generating functional for Green's functions should be
invariant under supersymmetry variations of the fields,
\begin{figure}[!ht]
\begin{center}
  \resizebox{10.5cm}{!}{\includegraphics{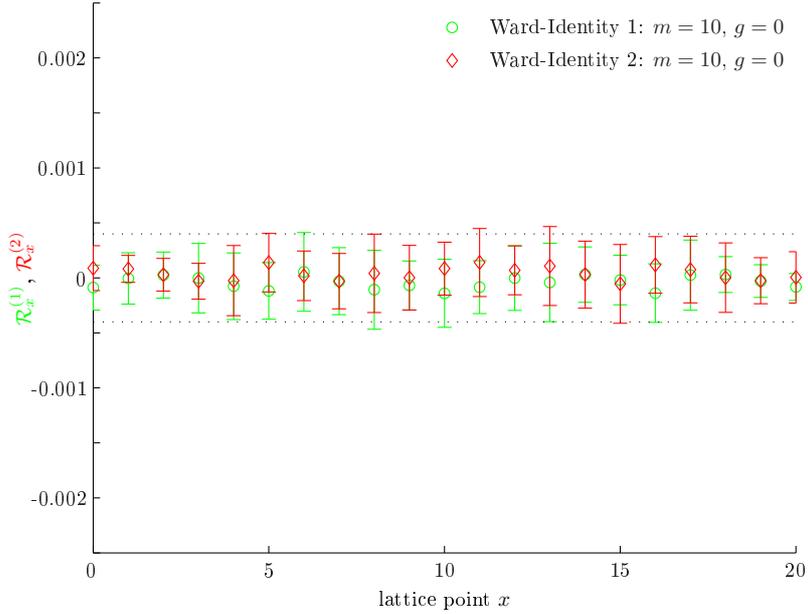}}
  \caption{Ward identities for the free theory, model~\textit{(ii)} (with shifted superpotential).}
  \label{fig:wi2f}
\end{center}
\end{figure}
\begin{equation}
  0 = \delta Z[J,\theta,\bar{\theta}] = \int {\cal D}(\phi,\psi)\,e^{-S + \sum_x (J_x\phi_x
      + \theta_x\psi_x + \bar{\theta}_x\psb_x)}\Big(\sum_y (J_y\delta\phi_y + \theta_y\delta\psi_y
      + \bar{\theta}_y\delta\psb_y) - \delta S\Big).
\end{equation}
Ward identities are obtained from derivatives of this equation with
respect to the sources. The second derivative $\pa^2/\pa J_x\pa\bar{\theta}_y$
leads to the Schwinger-Dyson equation
\eqnl{
  \langle\phi_x\psb_y\,\delta^{(1)} S\rangle = \langle\phi_x\,
    \delta^{(1)}\psb_y\rangle + \langle\psb_y\delta^{(1)}\phi_x\rangle,}{ward1}
whereas $\pa^2/\pa J_x\pa\theta_y$ yields
\eqnl{
  \langle\phi_x\psi_y\,\delta^{(2)} S\rangle = \langle\phi_x\,\delta^{(2)}
    \psi_y\rangle + \langle\psi_y\delta^{(2)}\phi_x\rangle.}{ward2}
For the improved models  \textit{(iv)}--\textit{(vi)} in
subsection \ref{sec:detail2} the actions are manifestly invariant 
under $\delta^{(1)}$, and the left-hand side of~\refs{ward1} vanishes.
Thus for these models the following Ward identities hold on
the lattice:
\eqnl{
  \langle\psi_x\psb_y\rangle - \langle\phi_x\xi_y\rangle = 0.}{ward3}
This can be confirmed in numerical checks and merely serves as a test bed for
the precision of the algorithms. The discretization of the second continuum
supersymmetry transformation, however, only leaves these lattice actions
invariant in the free case. With interactions, the term $\delta^{(2)} S$
leads to a nonvanishing left-hand side in the Schwinger-Dyson identities~%
\refs{ward2} which therefore measures the amount by which the second continuum
supersymmetry is broken by the discretization. Since this supersymmetry can
be made manifest by choosing different Nicolai variables $\tilde \xi$, the
supersymmetry breaking terms are the difference of both actions~\refs{impact5}
and~\refs{impact11}, i.\,e., the discretization of the surface terms (for
\textit{(iv)} and \textit{(vi)}).

For the actions in \textit{(i)}--\textit{(iii)} all supersymmetries are broken
by the discretization; the corresponding Schwinger-Dyson equations again 
can serve as a measure for the quality of the lattice approximation to supersymmetry.
Barring interactions, we expect them to hold for both supersymmetry transformations
in \textit{(ii)} and \textit{(iii)}.
\begin{figure}[!t]
\begin{center}
  \resizebox{10.5cm}{!}{\includegraphics{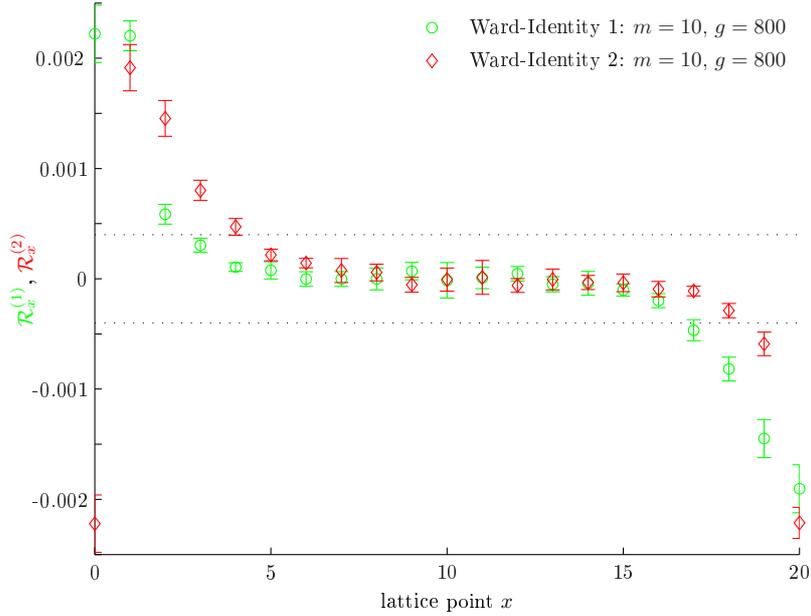}}
  \caption{Ward identities for the theory at strong coupling, model~\textit{(ii)} (with shifted superpotential).}
  \label{fig:wi2i}
\end{center}
\end{figure}

\subsubsection{Ward identities of the unimproved models}
The Ward identities have been simulated for models with the same superpotential as
in~\refs{intm1}. Since at weak coupling similar results can be expected as in the
free theory with $g=0$, we have also studied the theory at strong
coupling with $g=800$ and $m=10$ corresponding to a dimensionless ratio $g/m^2=8$.

The naive discretization~\textit{(i)} is not supersymmetric even without interactions,
therefore we concentrate on the other models where the supersymmetry of the free
theory sets a scale for the quality of the simulation of the broken supersymmetries
in the interacting case. For the unimproved models, we can use translation invariance
and measure the right-hand sides of the Ward identities~\refs{ward1} and~\refs{ward2},
i.\,e.
\eqnl{
  \mathcal{R}^{(1)}_{x-y} = \langle\psi_x\psb_y\rangle - \langle\phi_x (\pa\phi)_y\rangle -
  \langle\phi_x W_y\rangle}{wardui1}
and
\eqnl{
  \mathcal{R}^{(2)}_{x-y} = \langle\phi_x\,(\pa\phi)_y\rangle - \langle\phi_x W_y\rangle -
  \langle\psb_x\psi_y\rangle}{wardui3}
as functions of $x-y$. It should be noted that without interactions, the first Ward
identity reduces to a matrix identity $D^{-1} - (D^T D)^{-1} D^T=0$ for the free
Dirac operator $D_{xy}=\pa_{xy} + m\delta_{xy}$ if one uses that $\langle\psi_x
\psb_y\rangle=D^{-1}_{xy}$ and $\langle\phi_x\phi_y\rangle=(D^T D)^{-1}_{xy}$.
The corresponding data for the free theory in the case of the model with shifted
superpotential (model~\textit{(ii)}) are shown in figure~\ref{fig:wi2f}. In order to
keep statistical errors small, in all situations $4$~runs with
$10^6$~independent configurations were evaluated.
Within our numerical precision, supersymmetry is broken for this model if the simulation
data in the interacting case exceeds the bounds set by the free theory. The results
for~\refs{wardui1} and~\refs{wardui3} at $g=800$ are displayed in figure~\ref{fig:wi2i}.
Remarkably, the statistical error is much smaller than in the free theory; supersymmetry
breaking is equally strong for both supersymmetries~$\delta^{(1)}$ and~$\delta^{(2)}$ from~\refs{lsqm13} at strong coupling.
\begin{figure}[!t]
\begin{center}
  \resizebox{10.5cm}{!}{\includegraphics{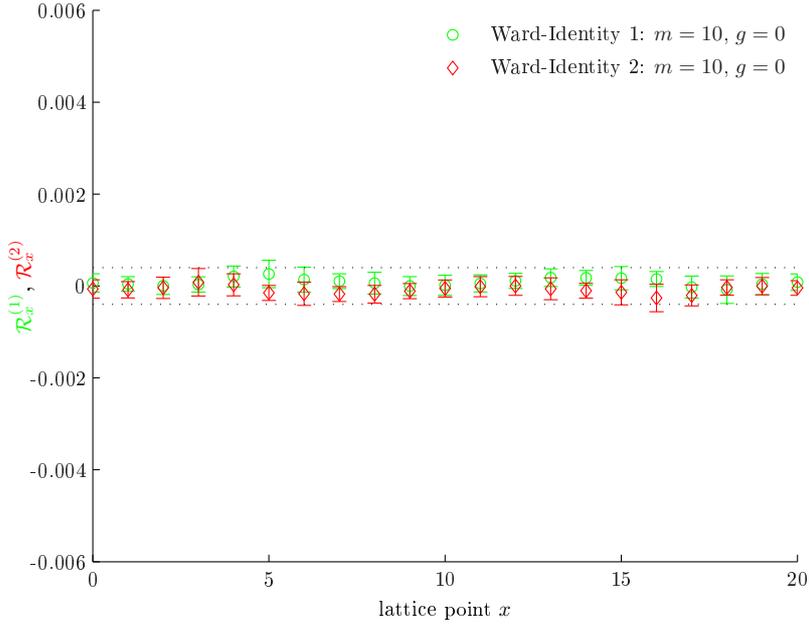}}
  \caption{Ward identities for the free theory, model~\textit{(iv)} (Wilson fermions with
  the Ito prescription).}
  \label{fig:wi4f}
\end{center}
\end{figure}

\subsubsection{Ward identities of the improved models}
\begin{figure}[!ht]
\begin{center}
  \resizebox{10.5cm}{!}{\includegraphics{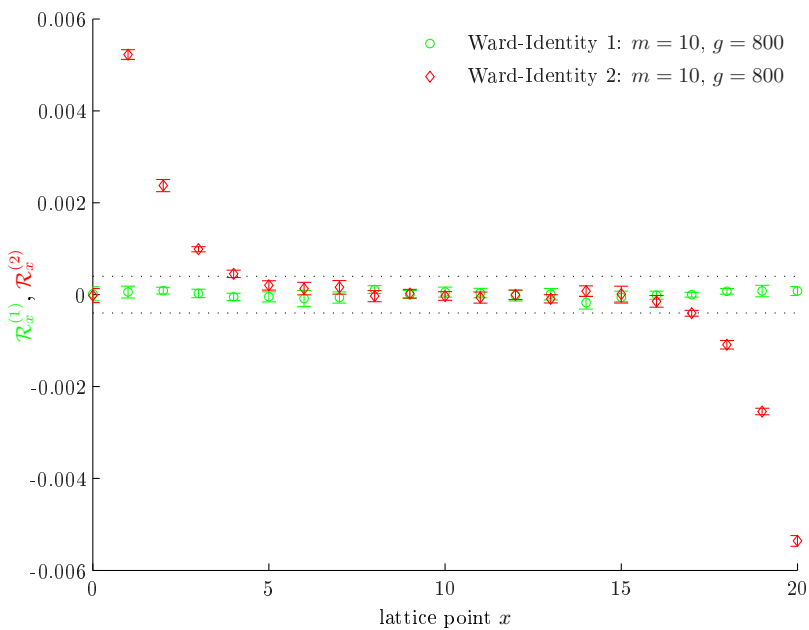}}
  \caption{Ward identities for the theory at strong coupling, model~\textit{(iv)}
  (Wilson fermions with the Ito prescription).}
  \label{fig:wi4i}
\end{center}
\end{figure}

For Wilson fermions with exact supersymmetry and the Ito prescription
(model~\textit{(iv)}), the statistical error as measured by the free
Ward identities is roughly of the same size as for the unimproved
model~\textit{(ii)}, cf.\ figure~\ref{fig:wi4f}. Again, this was obtained
by $4$~runs with $10^6$~independent configurations.
As expected, the first supersymmetry~$\delta^{(1)}$ (cf.~\refs{impact3} and
figure~\ref{fig:wi4i}) is preserved even at strong coupling; however, the
supersymmetry breaking effects for $\delta^{(2)}$ are about three
times as large as for the corresponding symmetry in the unimproved situation.

\begin{figure}[!p]
\begin{center}
  \resizebox{10.5cm}{!}{\includegraphics{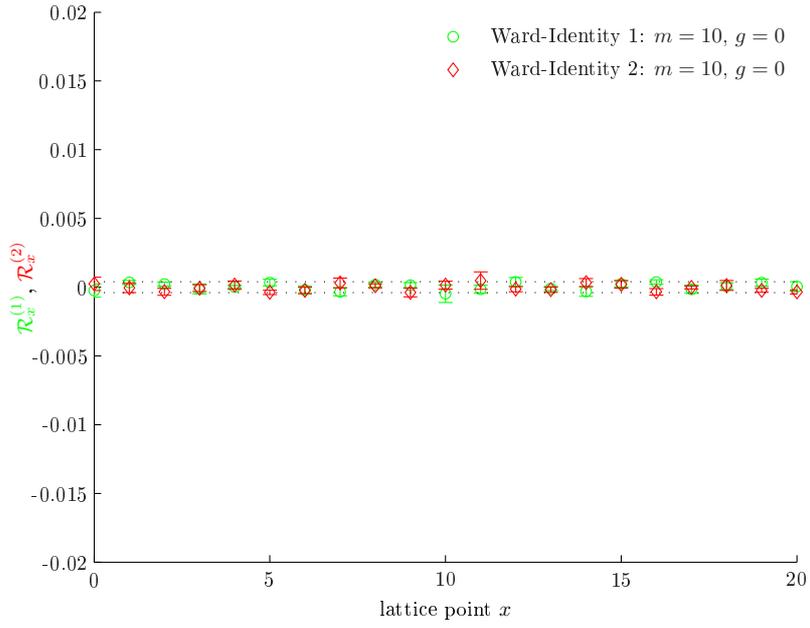}}
  \caption{Ward identities for the free theory, model~\textit{(v)}
  (Wilson fermions with the Stratonovitch prescription).}
  \label{fig:wi5f}
\end{center}
\end{figure}
\begin{figure}[!p]
\begin{center}
  \resizebox{10.5cm}{!}{\includegraphics{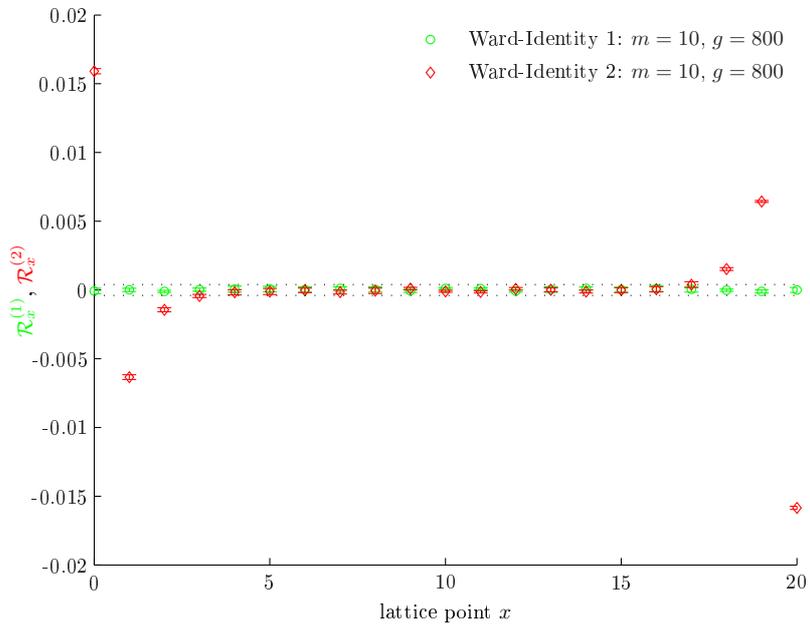}}
  \caption{Ward identities for the theory at strong coupling, model~\textit{(v)}
  (Wilson fermions with the Stratonovitch prescription).}
  \label{fig:wi5i}
\end{center}
\end{figure}
For the free supersymmetric model with Stratonovich prescription
(model~\textit{(iv)}), the second supersymmetry~\refs{impact9} is only a
symmetry with the definition~\refs{model5a} of $\tilde{\xi}$. The Ward
identities of both supersymmetry transformations,
\eqngrl{
  \langle\psi_x\psb_y\rangle - \langle\phi_x (\pa\phi)_y\rangle - \big\langle\phi_x\,
  W'\big(\ft{\phi_x+\phi_{x-1}}{2}\big)\big\rangle & = & 0,}{\langle\phi_x\,(\pa\phi)_y\rangle
  - \big\langle\phi_x\, W'\big(\ft{\phi_x+\phi_{x+1}}{2}\big)\big\rangle
  - \langle\psb_x\psi_y\rangle & = & 0,}{wardi1}
reduce in the free theory with $W'(\phi)=m\phi$ to matrix identities for
the free Dirac operator $D_{xy} = \pa_{xy} + \frac{m}{2}(\delta_{xy} + \delta_{x,y-1})$,
\eqngrl{
  D^{-1}-(D^T D)^{-1} D^T & = & 0,}{-(D^T D)^{-1}D + (D^{-1})^T & = & 0.}{wardi3}
Here, the second identity holds since $D$ is circulant and therefore normal.
The corresponding plot of the left-hand sides is shown in figure~\ref{fig:wi5f}.
This determines the mean error above which we take supersymmetry to be broken if we switch
on interactions. In figure~\ref{fig:wi5i}, the first supersymmetry is preserved within a
high numerical accuracy whereas the second supersymmetry is clearly broken by effects about
three times the size of the supersymmetry violation in the model with Ito
prescription.

For the supersymmetric model with SLAC derivative, the Ward identities are satisfied
within statistical error bounds, but determining the mean error in an
analogous manner fails since the contributions of the bosonic two-point functions
$\langle\phi_x(\pa\phi)_y\rangle$ in the free theory lead to large errors which
obscure the interpretation of the corresponding Ward identities.
\begin{figure}[!ht]
\begin{center}
  \resizebox{10.5cm}{!}{\includegraphics{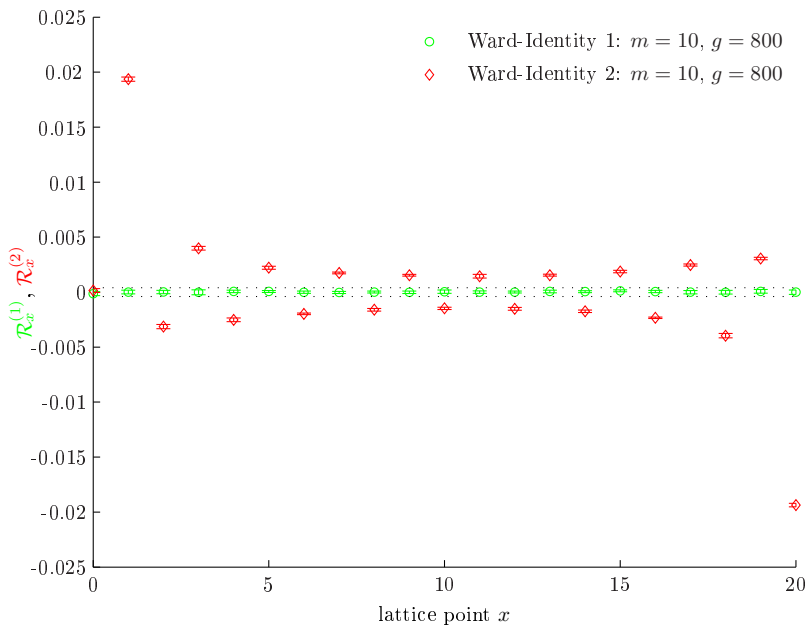}}
  \caption{Ward identities for the theory at strong coupling, model~\textit{(vi)}
  (supersymmetric model with SLAC fermions).}
  \label{fig:wi6i}
\end{center}
\end{figure}
At strong coupling, however, the interaction terms dominate the action, and the
fluctuations of the bosonic propagators become increasingly less important so that
we obtain rather precise results at $g=800$. Within the error bounds of
model~\textit{(v)}, the first supersymmetry is obviously preserved in the
interacting theory (cf. fig.~\ref{fig:wi6i}), whereas the supersymmetry breaking
effects of the second supersymmetry are about as large as for Wilson fermions with
the Stratonovich prescription.
\enlargethispage{3mm}
%%%%%%%%%%%%%%%%%%%%%%%%%%%%%%%%%%%%%%%%%%%%%%%

\section{The \texorpdfstring{${\cal N}=2$}{N=2} Wess-Zumino model in 2 dimensions}\label{sec:WZtwodim}
In this section, we study different discretizations of the two-dimensional
Wess-Zumino model with $(2,2)$ supersymmetry. The minimal variant 
contains a Dirac spinor field and two real scalar fields $\varphi_a$
which are combined into a complex scalar field $\phi=\varphi_1+i\varphi_2$. 
Also, we use the complex coordinate $z=x^1+ix^2$ and its complex conjugate
$\bar z$ in Euclidean spacetime and denote the corresponding derivatives
by $\pa=\half(\pa_1-i\pa_2)$ and $\bar\pa$, respectively. The Euclidean action 
contains the first and second derivatives $W'$ and $W''$ of a holomorphic
superpotential $W(\phi)$ with respect to the complex field $\phi$,
\eqnl{
S_{\rm cont} = 
\int d^2x\Big(
    2 \bar\pa\bar\phi \pa\phi + \half|W'|^2
+\bar\psi M\psi\Big),\quad
M=\slashed{\pa} +W''P_++\bar W''P_-,}{wz3}
where $P_\pm=\half(\mathbbm{1}\pm \gamma_3)$ are the chiral projectors.
In the Weyl basis with $\gamma^1 = \sigma_1$, $\gamma^2 = -\sigma_2$ 
and $\gamma_3 = i\gamma^1\gamma^2 = \sigma_3$, the complex spinors
can be decomposed according to 
\eqnl{
\psi=\pmatrix{\psi_1\cr \psi_2}\mtxt{and}
\psb=\big(\psb_1,\psb_2\big).}{wz5}
In this basis, the supersymmetry transformations
leaving $S_{\rm cont}$ invariant are
\begin{equation}
  \setlength{\arraycolsep}{2.5pt}
  \begin{array}{rl@{\qquad}rl@{\qquad}rl}
    \de\phi & = \psb^1\ve_1 + \veb_1\psi^1, & \de\psb^1 & = -\frac{1}{2}\Wb'\veb_1-\pa\phi\veb_2, &
      \de\psi^1 & = -\frac{1}{2}\Wb'\ve_1+\bar{\pa}\phi\ve_2,\\
    \de\phb & = \psb^2\ve_2 + \veb_2\psi^2, & \de\psb^2 & = -\bar{\pa}\phb\veb_1-\frac{1}{2}W'\veb_2, &
      \de\psi^2 & = \pa\phb\ve_1-\frac{1}{2}W'\ve_2.
  \end{array} \label{wz7}
\end{equation}
Similarly as in quantum mechanics a naive discretization of this model breaks
all four supersymmetries. In order to keep one supersymmetry
one can add an improvement term. In what follows we shall only consider
improved models; they differ by our choice of the lattice 
derivatives. Instead of trying to generalize the Stratonovich prescription 
to the two-dimensional situation, 
we find in subsection~\ref{sec:WZd2} that a
non-standard choice of the Wilson term leads to an improved behavior in the limit
of vanishing lattice spacing. This is corroborated by the results of our
simulations for the case of a cubic superpotential $W=\ft12 m\phi^2+\ft13 g\phi^3$
which we present in subsection~\ref{sec:WZres}.

\subsection{Lattice models with improvement}\label{sec:WZd2}
We start with a two-dimensional periodic $N_1\times N_2$ lattice $\Lambda$
with complex bosonic variables $\phi_x$ and two complex spinors
$\psi_x,\psb_x$ on the lattice sites $x=(x^1,x^2)\in \Lambda$. 
Again two different \emph{antisymmetric} lattice derivatives in direction
$\mu$ are used. These are the ultralocal derivative
\eqnl{
\pah_{\mu,xy}=\frac{1}{2a}\left(\delta_{x+e_\mu,y}-\delta_{x-e_\mu,y}\right)}{wz11}
with doublers and the nonlocal SLAC derivative without doublers, which
for odd $N_1,N_2$ reads
\eqnl{
\paslac_{1,xy}=\paslac_{x_1 \neq y_1}\delta_{x_2,y_2},\quad
\paslac_{2,xy}=\paslac_{x_2\neq y_2}\delta_{x_1,y_1}\mtxt{and}
\paslac_{\mu,xx}=0}{wz13}
(analogously to the one-dimensional SLAC derivative defined in~\refs{latder3}).
Later we shall remove the fermionic doublers of $\gamma^\mu\pah_\mu$ 
by introducing two types of Wilson terms, both containing 
the lattice Laplacian $\Delta$ defined by
\eqnl{
(\Delta \phi)(x)=\frac{1}{a^2}\Big(\sum_{\mu=1,2}\big[\phi(x+e_\mu)
+\phi(x-e_\mu)\big]-4\phi(x)\Big).}{wz15}
As for the continuum model we use the holomorphic lattice derivative 
$\pa_{xy} := \half(\pa_{1,xy} - i \pa_{2,xy})$.

The Nicolai variables of one-dimensional systems are easily generalized 
to two dimensions,
\eqnl{
  \xi_x = 2(\bar\pa\bar\phi)_x + W_x,\qquad
\bar{\xi}_x =2(\pa\phi)_x + \Wb_x;}{wz17}
again, $W_x$ denotes terms (to be specified below) derived from the
superpotential. The bosonic part of the action is Gaussian in these
variables,
$S_{\rm bos}=\frac{a^2}{2}\sum_{x} \bar{\xi}_x \xi_x$,
and has the explicit form
\eqnl{
S_{\rm bos} = 
a^2\sum_{x}\Big(
    2 (\bar\pa\bar\phi)_x (\pa\phi)_x + W_x(\pa\phi)_x
    + \Wb_x(\bar\pa\bar{\phi})_x + \half|W_x|^2\Big).}
{wz19}
For antisymmetric derivatives $\pa_\mu$, the kinetic term has
the standard form $\sum_{a,\mu=1}^2 (\pa_\mu \varphi_a)_x^2$ in terms of the real
fields $\varphi_a$; in particular this holds true for the ultralocal derivative $\pah_\mu$ 
and the SLAC derivative $\paslac_\mu$ introduced above. The second and third term
in $S_{\rm bos}$ are absent in the continuum action.
In a naive discretization of the continuum model these
\emph{improvement terms} do not show up. In the continuum limit
they become surface terms and could be dropped. On the
lattice they are needed to keep one of the four supersymmetries 
intact.

Supersymmetry requires an additional fermionic term
$ S_{\rm ferm} = a^2\sum\psb_x M_{xy}\psi_y$
for a two-component Dirac spinor field in such a way that the determinant
of the Jacobian matrix
\eqnl{
%\frac{\pa(\xi,\bar\xi)}{\pa(\phi,\bar\phi)}=
  \pmatrix{
    \pa\xi_x/\pa\phi_y & \pa\xi_x/\pa\bar{\phi}_y \cr
    \pa\bar{\xi}_x/\pa\phi_y & \pa\bar{\xi}_x/\pa\bar{\phi}_y}
  = \pmatrix{
    W_{xy} & 2\bar\pa_{xy}\cr
    2\pa_{xy} & \Wb_{xy}},\quad 
W_{xy}=\frac{\pa W_x}{\pa\phi_y},}{wz21}
for the change of bosonic variables $(\phi,\bar{\phi})\mapsto(\xi,\bar{\xi})$
cancels the fermion determinant $\det M$. Actually the fermionic operator
$M$ in \refs{wz3} with $\gamma$-matrices in the Weyl basis and continuum
derivatives replaced by lattice derivatives is identical 
to the Jacobian matrix. Hence we choose as the fermionic part of
the action
\eqngrl{
S_{\rm ferm}=a^2\sum_{x,y} \psb_x M_{xy}\psi_y,&&
M_{xy}=\gamma^\mu \pa_{\mu,xy}+W_{xy}P_+
+\bar W_{xy}P_-}
{&&\hskip9mm =M_0+
 W''(\phi_x)\delta_{xy}\,P_+ + \bar W''(\phi_x)\delta_{xy}\,P_-\,.}
{wz25}
By construction the action $S_{\rm susy}=S_{\rm bos}+S_{\rm ferm}$ 
with improvement terms is invariant under the supersymmetry
transformations generated by $\delta^{(1)}$:
\begin{equation}
  \setlength{\arraycolsep}{2.5pt}
  \begin{array}{rl@{\qquad}rl@{\qquad}rl}
    \de^{(1)}\phi_x & = \veb\psi^1_x, & \de^{(1)}\psb^1_x & = -\frac{1}{2}\veb\bar{\xi}_x,
      & \de^{(1)}\psi^1_x & = 0,\\
    \de^{(1)}\phb_x & = \veb\psi^2_x, & \de^{(1)}\psb^2_x & = -\frac{1}{2}\veb\xi_x,
      & \de^{(1)}\psi^2_x & = 0;
  \end{array}
\end{equation}
this corresponds to a discretization of the continuum symmetry~\refs{wz7} with
$\ve_1=\ve_2=0$ and $\veb_1=\veb_2=\veb$. The other three continuum 
supersymmetries are broken; for appropriately chosen Nicolai variables $\xi$ an
action preserving any one of the three other supersymmetries can 
be constructed analogously~\cite{CatterallKaramov2}.\footnote{The corresponding
Nicolai variables can be read off from the right-hand sides of $\de\psi^a$
in~\refs{wz7} for $\ve_1=\pm\ve_2=\ve$ and $\veb=0$ or from the right-hand
sides of $\de\psb^a$ for $\veb_1=\pm\veb_2=\veb$ and $\ve=0$.} In this paper,
we are going to use the Nicolai variable \refs{wz17} and consider
several possibilities to remove fermion doublers.
%%%%%%%%%%%%%%%%%%%%
\subsection{The lattice models in detail}\label{sec:tdlatmod}
We introduce three different lattice approximations to the
continuum Wess-Zumino model \refs{wz3}.
They are all equipped with an improvement term and thus admit 
one supersymmetry. The first two models contain Wilson fermions and the 
third the SLAC derivative. It will turn out that the discretization
errors of the eigenvalues of the bosonic and fermionic kinetic operators
in the free case indicate how good the approximation to the continuum
theory is when we turn on interactions.
\\[2mm]
\emph{(i) Supersymmetric model with standard Wilson term}\nopagebreak \\ \nopagebreak
Here we choose ultralocal derivative $\pah_\mu$ and add a standard
Wilson term to the superpotential to get rid of the doublers of
$\gamma^\mu\pah_\mu$ so that
\eqnl{
S^{(1)}=S_{\rm bos}+S_{\rm ferm}\mtxt{with}
\pa_\mu=\pah_\mu,\quad W_x=-\frac{ar}{2}(\Delta\phi)_x+W'(\phi_x).}{stand1}
For later convenience we do not fix the Wilson parameter $r$ in
this section. In this lattice model the Dirac operator $M_0$ in~\refs{wz25}
takes the form
\eqnl{
M^{(1)}_0=\gamma^\mu \pah_\mu-\frac{ar}{2}\Delta
}{stand3}
and we easily recognize the standard Dirac operator for Wilson fermions.
The bosonic part of the action may be expanded as
\eqnl{
S^{(1)}_{\rm bos} =\frac{a^2}{2}\sum_{x} \Big(\bar\phi_x (K\phi)_x
+ |W'(\phi_x)|^2\Big)
+a^2\sum_{x}\Big(W'(\phi_x)\,\big(\ring{\pa}\phi
-\frac{ar}{4}\Delta\bar\phi\big)_x + \hbox{c.c.}\Big)  
}{stand5}
with the kinetic operator
\eqnl{
K =-\ring{\Delta}+\big(\ft12 ar \Delta\big)^2,\mtxt{where}
\ring{\Delta}=-\pah_\mu\pah^\mu.}{stand7}
The last term in \refs{stand5} is the improvement term --- a discretization
of a surface term in the continuum theory. Note that even for the
\emph{free massive model} with $W'(\phi)=m\phi$ the improvement 
term $-\half (amr)\,a^2\sum \bar\phi_x(\Delta\phi)_x$ is non-zero
such that \refs{stand5} becomes
\eqnl{
S^{(1)}_{\rm bos}=\frac{a^2}{2}\sum_{xy} \bar\phi_x K^{(1)}_{xy}\phi_y\mtxt{with}
K^{(1)}=K+m^2-arm\Delta.}{stand9}
The eigenvalues of the commuting operators $-\ring{\Delta}$
and $-\Delta$ are $\ring{p}^2$ and $\hat{p}^2$, where
\eqnl{
\accentset{\circ}{p}_\mu=\frac{1}{a}\sin a p_\mu,\quad 
\hat{p}_\mu=\frac{2}{a}\sin\Big(\frac{a p_\mu}{2}\Big)\mtxt{with}
p_\mu=\frac{2\pi k_\mu}{a N_\mu},\quad k_\mu\in\{1,2,\dots,N_\mu\},}{stand11}
such that the eigenvalues of the matrix $K^{(1)}$ in \refs{stand9} are
\eqnl{
  \mu_p = \accentset{\circ}{p}^2 + \big(m+\ft12 ar \hat{p}^2\big)^2.}{stand13}
On the other hand, the eigenvalues of the free Dirac operator $M^{(1)}_0+m$
are given by 
\eqnl{
\l_p^\pm = m+\ft12 ar \hat{p}^2\pm i|\accentset{\circ}{p}|.}{stand15} 
Thus, $\mu_p=\l_p^+ \l_p^-$, and the fermionic and bosonic determinants coincide
for the free theory. The discretization errors for these eigenvalues are of order $a$,
\eqnl{
  \mu_p  = p^2 + m^2 + (ar m)  p^2 + O(a^2),\quad
 \l_p^\pm = \pm i\vert p\vert +m+\ft12 ar\, p^2+O(a^2).}{stand17}
It is remarkable that the bosonic part of the action in the continuum is an even
function of the mass, whereas its discretization~\refs{stand5} is not. This
spoils the sign freedom in the fermion mass term on the lattice and motivates
the following second possibility.\\[2mm]
%%%%%%%%%%%%%%%%%%%%%%%%%%%%
\emph{(ii) Supersymmetric model with non-standard Wilson term}\\
Again we choose the ultralocal derivatives $\pah_\mu$ but now add a
non-standard Wilson term to the superpotential to get rid of the
doublers, so that the action now is
\eqnl{
S^{(2)}=S_{\rm bos}+S_{\rm ferm}\mtxt{with}
\pa_\mu=\pah_\mu,\quad W_x=\frac{iar}{2}(\Delta\phi)_x+W'(\phi_x).}{twist1}
This choice is equivalent to an imaginary value of the Wilson parameter
inside the holomorphic superpotential. Now the Dirac operator $M_0$
in~\refs{wz25} has the form
\eqnl{
M_0^{(2)}=\gamma^\mu\pah_\mu+\frac{iar}{2}\gamma_3\Delta}{twist3}
and contains a non-standard Wilson term, reminiscent of a momentum
dependent twisted mass. It should be noted that the only difference
between the bosonic actions~\refs{stand5} and
\eqnl{
S^{(2)}_{\rm bos} =\frac{a^2}{2}\sum_{x} \Big(\bar\phi_x (K\phi)_x+
 |W'(\phi_x)|^2\Big)
+a^2\sum_{x}\Big(W'(\phi_x)\,\big(\ring{\pa}\phi-\frac{iar}{4}\Delta\bar\phi\big)_x 
+ \hbox{c.c.}\Big),}{twist5}
is the improvement term.
The modified Wilson term in \refs{twist1} yields an action which is even in the mass $m$.
Actually, for the \emph{free massive model} with $W'(\phi)=m\phi$ the improvement term
vanishes and
\eqnl{
S^{(2)}_{\rm bos}=\frac{a^2}{2}\sum_{xy} \bar\phi_x K^{(2)}_{xy}\phi_y\mtxt{with}
K^{(2)}=K+m^2.}{twist7}
The eigenvalues of $K^{(2)}$ and of the free Dirac operator $M^{(2)}+m$
in this case are given by 
\eqnl{
\mu_p=m^2+\accentset{\circ}{p}^2+\left(\ft12 ar\,\hat{p}^2\right)^2\mtxt{and}
\l_p^\pm = m\pm i\sqrt{\accentset{\circ}{p}^2 + \left(\ft12 ar\hat{p}^2\right)^2}.}{twist9}
Again the determinants of fermionic and bosonic operators are equal. The added
advantage in this situation is, however, that the discretization errors of the
continuum eigenvalues are only of order $O(a^2)$. Namely, for small lattice spacing,
\eqnl{
\mu_p=m^2 + p^2 + \kappa a^2 + O(a^4),\quad 
\l_p^\pm= m\pm i(p^2+\kappa a^2)^{1/2}+ O(a^4),}{twist13}
where the $O(a^2)$-coefficient
\eqnl{\kappa = -\frac{1}{3}\sum_\mu p_\mu^4 + \frac{r^2}{4}\Big(\sum_\mu p_\mu^2\Big)^2
             = \Big(\frac{r^2}{4}-\frac{1}{3}\Big)\sum_\mu p_\mu^2
                 + \frac{r^2}{2}(p_1)^2(p_2)^2}{twist14}
vanishes for $3r^2=4$ and $p_1=0$ or $p_2=0$. As explained in section~\ref{sec:filter},
we take spatial averages over two-point functions for the mass extraction; this
projects the spatial momentum to zero. Thus, one might expect discretization
errors of order $O(a^4)$ for
\begin{equation}
  r^2 = \frac{4}{3}. \label{twist15}
\end{equation}
In fact we will see in section~\ref{sec:WZres} that this choice
leads to the best continuum approximation -- and this in spite of the fact that
it violates reflection positivity (as does the improvement
term in all supersymmetric models).\\[2mm]
%%%%%%%%%%%%%%%%%%%%%%%%%%%%%%%%%%%%%%%
\emph{(iii) Supersymmetric model with SLAC derivative}\\
The Dirac operator $\gamma^\mu\paslac_\mu$ with nonlocal and antisymmetric 
SLAC derivatives defined in~\refs{wz13} has no doublers and no Wilson terms
are required; this leads to the action
\eqnl{
S^{(3)}=S_{\rm bos}+S_{\rm ferm}\mtxt{with}
\pa_\mu=\paslac_\mu,\quad W_x=W'(\phi_x)}{slac1}
with Dirac operator
\eqnl{
%M^{(3)}=M_{0,xy}^{(3)}+W''(\phi_x)\delta_{xy}P_+
%+\bar W''(\phi_x)\delta_{xy}P_- \mtxt{with}
M_0^{(3)}=\gamma^\mu\paslac_\mu,}{slac3}
cf.~\refs{wz25}. For the free massive model with SLAC
derivative the improvement term vanishes, in particular the bosonic part of
$S^{(3)}$ is
\eqnl{S^{(3)}_{\rm bos}=\frac{a^2}{2}\sum_{xy} \bar\phi_x K^{(3)}_{xy}\phi_y
  \mtxt{with}K^{(3)}=-\Delta^{\rm slac}+m^2;}{slac5}
all supersymmetries are realized. We shall see in section~\ref{sec:WZres}
that $S^{(3)}$ is a very good approximation to the continuum model
and in section~\ref{sec:three} that the lattice model based on the
SLAC derivative is one-loop renormalizable in spite of its nonlocality.
%%%%%%%%%%%%%%%%%%%%%%%%%%%%%%
\subsection{Simulations of the Wess-Zumino model}\label{sec:WZres}
In subsection~\ref{sec:WZd2}, we have formulated the model in a complex basis,
which is natural and convenient for models with two supersymmetries (in
particular, the simplest form of the Nicolai map~\refs{wz17} is in terms
of the complex scalar fields $\phi=\varphi_1+i\varphi_2$ and
$\xi=\xi_1+i\xi_2$). On the other hand, for numerical simulations it is
convenient to have a formulation of the model in terms of the real
components $\varphi_a$ and $\xi_a$ which are combined to real doublets,
\eqnl{
\mbvarphi=\pmatrix{\varphi_1\cr \varphi_2}\mtxt{and}
\mbxi=\pmatrix{\xi_1\cr\xi_2}.}{sim2d1}
As to the fermions, it is most appropriate to use a Majorana representation
with real $\gamma$-matrices $\gamma^1=\sigma_3,\;\gamma^2=\sigma_1$
such that $\gamma_3=i\gamma^1\gamma^2=-\sigma_2$.
All simulations were done for the model with cubic superpotential
$W=\ft12 m\phi^2+\ft13 g\phi^3$ with derivative
\eqnl{
W'(\phi)=
m\varphi_1+u+i(m\varphi_2+v),}{sim2d3}
where we have introduced the abbreviations $u=g(\varphi_1^2-\varphi_2^2)$ and
$v=2g\varphi_1\varphi_2$. The actions contain a quartic potential
\eqnl{
V(\mbvarphi)=(g^2 \mbvarphi^2+2mg\varphi_1)\,\mbvarphi^2.}{sim2d5}
In terms of real fields, the Nicolai map \refs{wz17} takes the form
\eqnl{
\mbxi^{(n)}=(M^{(n)}_0+m)\mbvarphi+\pmatrix{u\cr v}
\mtxt{for}n=1,2,3}{sim2d7}
with model-dependent free massless Dirac operators $M^{(n)}_0$
as given in~\refs{stand3}, \refs{twist3} and~\refs{slac3}.
%%%%%%%%%%%%%%%%%%%%%%%%%%%
The bosonic actions for the models with standard Wilson, modified Wilson
and SLAC fermions can now be written as
\eqnl{
S^{(n)}_{\rm bos}=\frac{a^2}{2}\sum_{xy} (\mbvarphi_x,K^{(n)}_{xy}\mbvarphi_y)+
\frac{a^2}{2}\sum_x V(\mbvarphi)+\Delta^{(n)}}
{sim2d9}
with model-dependent kinetic operators $K^{(n)}$ as introduced in~\refs{stand9},
\refs{twist7} and~\refs{slac5}. In terms of the divergence and curl of a vector
field in two dimensions, $\hbox{div }\mbvarphi=\pa_1 \varphi_1+\pa_2\varphi_2$
and $\hbox{curl }\mbvarphi=\pa_1\varphi_2-\pa_2\varphi_1$, the model-dependent
improvement terms
\begin{eqnarray}
\Delta^{(1)} & = & a^2\sum_x u_x\left((\hbox{div }\mbvarphi)_x - \frac{a r}{2}
  (\Delta\varphi_1)_x\right) - a^2\sum_x v_x \left((\hbox{curl }\mbvarphi)_x +
  \frac{a r}{2}(\Delta \varphi_2)_x\right),\nonumber\\
\Delta^{(2)} & = & a^2\sum_x u_x\left((\hbox{div }\mbvarphi)_x - \frac{a r}{2}
  (\Delta\varphi_2)_x\right) + a^2\sum_x v_x \left(-(\hbox{curl }\mbvarphi)_x +
  \frac{a r}{2}(\Delta \varphi_1)_x\right),\label{sim2d11}\\
\Delta^{(3)} & = & a^2\sum_x u_x(\hbox{div }\mbvarphi)_x - a^2\sum_x v_x
  (\hbox{curl }\mbvarphi)_x\nonumber
\end{eqnarray}
again are discretizations of continuum surface terms.

\subsubsection{Models without interaction}
As expected, the masses of all models without interactions converge to the same
continuum limit~$m$. Since all free models are supersymmetric (w.\,r.\,t.\ two
supersymmetries), boson and fermion masses extracted from the two-point functions
coincide even at finite lattice spacing.
The masses $m(a)$ for the models \textit{(ii)} and \textit{(iii)} with non-standard
Wilson term (with $r^2=4/3$) and the SLAC derivative, respectively, at finite~$a$ are already
very close to their continuum limits, the effective mass as a function of $a$ for the
model with standard Wilson term (with $r=1$) has a much larger slope.\footnote{This
behavior is reminiscent of the masses $m(a)$ for the corresponding quantum mechanical
model in section~\refs{sec:massesqmfree}.} This is in line with the approximation of
the eigenvalues \refs{stand17} and \refs{twist13} to those of the continuum kinetic
operators.

\subsubsection{Models with interaction}\label{sec:WZmasswint}
We have calculated the masses for the interacting models with cubic superpotential
\eqnl{W(\phi)=\frac{1}{2}m\phi^2 + \frac{1}{3}g\phi^3}{d2intm1}
for masses $m=10$ and couplings $g$ ranging between $0$ and $1$. 
The effective masses at different values of the lattice spacing are determined as described in section \ref{sec:filter}.
Again, due to
supersymmetry boson and fermion masses coincide also for finite lattice spacing. In
all cases, they converge to a continuum value which however cannot reliably be
distinguished from the value $m_{\rm free}(a=0)=10$ within error bounds.
This is to be expected since in continuum perturbation theory, the one-loop
corrected mass is
\eqnl{m_{1-{\rm loop}} = m\Big(1-\frac{g^2}{4\pi m^2}\Big)}{d2intm3}
in a renormalization scheme without wave-function renormalization (this corresponds
to a correction less than about $0.3\%$ with our values of $m$ and $g$).
Thus, significant effects should only be seen at larger values of $g/m$.
Unfortunately, our simulations require reweightings which lead to rather
large error bounds (which grow as the coupling increases). The convergence
behavior to the expected continuum value is model-dependent:
The model with the non-standard Wilson term shows the expected improved
behavior leading to good estimates for the continuum mass already at
finite lattice sizes. As in quantum mechanics, this applies also to
the SLAC derivative.

\begin{table}[!ht]
\begin{center}
\begin{tabular}{|c|c|c|c|c|}
\hline
 Lattice spacing & \multicolumn{2}{c|}{Wilson} & \multicolumn{2}{c|}{Twisted
 Wilson} \\ \cline{2-5}
 & $m_B$ & $m_F$ & $m_B$ & $m_F$\\
\hline
0.1250& 6.34(4)& 6.4872(1) & 9.51(8) & 9.95(7)\\
0.0833& 7.32(3)& 7.2730(2) & 10.0(1) & 10.0182(1)\\
 0.0625& 7.81(8)& 7.768(1) & 10.4(1) & 9.99(1)\\
 0.0500& 8.16(4)&  8.07(1) & 9.82(1) & 9.93(3)\\
\hline
\end{tabular}
\vskip3mm
\begin{tabular}{|c|c|c|}
 \hline
 Lattice spacing & \multicolumn{2}{c|}{SLAC}\\ \cline{2-3}
& $m_B$ & $m_F$\\
\hline
 0.0769& 9.9(2) & 10.0(2)\\
 0.0667& 10.0(1) & 10.0(1)\\
0.0526& 10.0(1) & 10.00(5)\\
 0.0400& 9.95(6) & 9.99(2)\\
 0.0323& 10.03(3) & 9.98(1)\\
 0.0213& 9.83(3) & 9.98(1) \\
\hline
\end{tabular}
\end{center}
\vskip-3mm
\caption{Comparison of extracted masses at $g=0.5$}
\end{table}
\vskip8mm
\begin{figure}[!h]
\begin{center}
  \resizebox{10.5cm}{!}{\includegraphics{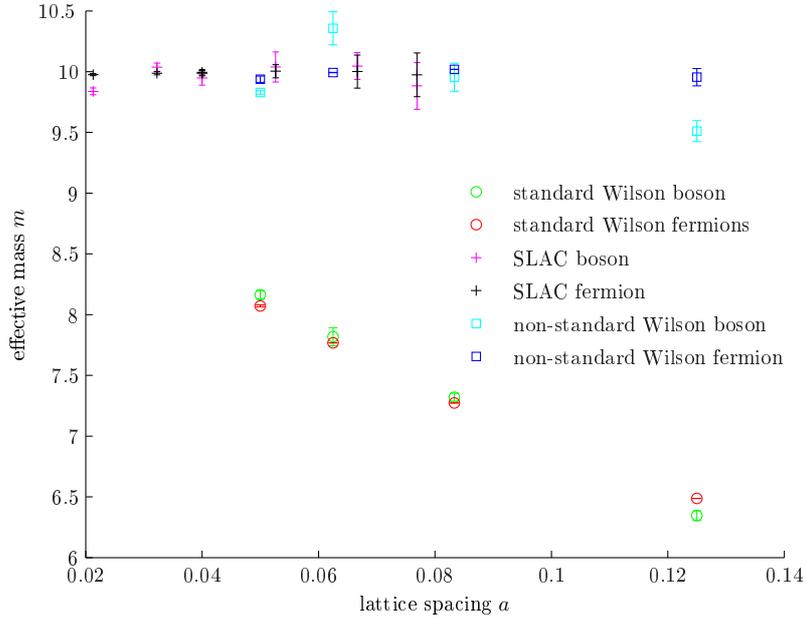}}
  \caption{Effective mass for the two-dimensional Wess-Zumino model at $g=0.5$}
  \label{fig:16}
\end{center}
\end{figure}
%%%%%%%%%%%%%%%%%%%%%%%%%%%%%%%%%%%%%%%%%%%%%%%%%%%%%%

\section{Algorithmic aspects}
\label{sec:four}
In this section we briefly outline the methods and algorithms we have used in
our simulations. In particular we have modified the treatment of the fermion
determinant in the well known Hybrid Monte Carlo algorithm (HMC)~\cite{Duane:1987de}.

Although due to the low dimensionality of the models under consideration
the numerical studies to be carried out are much less demanding than, e.\,g.,
four-dimensional lattice QCD we have to face similar problems
with respect to the treatment of the fermion fields. However, the computational
tasks at hand allow for strategies which are more accurate and
easier to implement than what is widely used there. Since our models
do not contain any gauge degrees of freedom, the resulting Dirac operators
have a rather simple form, a fact which can be put to practical use as will
be shown below.

Nonetheless, the mere presence of the fermion determinant in the partition
function introduces a nonlocality which has to be taken into account in the
Monte Carlo algorithm. We thus decided to base our numerical simulations on
the HMC which allows for a simultaneous update of all
bosonic field variables. For the quantum mechanical models discussed
in section \ref{sec:two} substantial improvements can be achieved if
it is possible to compute the fermion determinant in closed form. For the
two-dimensional models, however, comparable results are not available, and
these theories are plagued by a strongly fluctuating fermion determinant.
Even worse, the fermion determinant may take on positive and negative values
which drives the simulations directly into the so-called sign problem. To
proceed we will therefore treat the quantum mechanical models again separately
from the two-dimensional models and discuss them one at a time.

\subsection{Quantum Mechanics}
\label{subsec:algoQM}
Our setup for the bosonic degrees of freedom for the HMC does not
differ from the standard procedure and is formulated on an enlarged phase space
involving the real bosonic field $\phi_x$ and an additional conjugate momentum
field $\pi_x$. As usual these fields are propagated along the molecular dynamics
trajectory by Hamilton's equations
\begin{equation}
\label{eq:hamEqs}
\dot\phi_x=\frac{\partial H}{\partial \pi_x},\;\;\dot\pi_x=-\frac{\partial H}{\partial \phi_x},
\end{equation}
 where \begin{equation}
\label{eq:hamDefinition}
H=\frac{1}{2}\sum_x \pi_x^2+S[\phi].
\end{equation}
Since on the lattice the fermions are already integrated out the expression to
be used in eq.~\refs{eq:hamDefinition} is given by 
\begin{equation}
\label{eq:actionEff}
S[\phi]=S_B[\phi]+\ln\det M[\phi].
\end{equation}
In the standard approach one would now introduce a pseudofermion field $\chi$ to
obtain a stochastic estimate for $\det M[\phi]$ which however will necessarily
introduce additional noise to later measurements. Hence it would be clearly favorable
to take the fermion determinant exactly into account. While a direct
computation of the fermion determinant at each step of a trajectory is also
feasible in these one-dimensional theories one can do even better due to the
simple structure of the Dirac operator. Despite their differences in the bosonic
part of the action the models involving an antisymmetric derivative matrix and
a Wilson term, namely the models \textit{(i)}, \textit{(ii)}, and \textit{(iv)},
share the same fermion Matrix $M_W[\phi]$. For these cases a closed expression
for the determinant with Wilson parameter $r=1$ is given by~\refs{det5},
\begin{equation}
\label{eq:wilsonDet}
\det M_W[\phi] = \prod_x \left( 1+m+3g\phi_x^2\right) - 1.
\end{equation}
It should be noted that with this `diagonal' nonlocal form of the determinant even
local algorithms could have been used. A similar expression for the Stratonovich
prescription as given in~\refs{det11} also allows for a quick computation of the
fermion determinant in model~\textit{(v)}. For the SLAC derivative, it is easier
to again exploit the local structure of the interaction terms in the fermion
matrix by varying 
\begin{equation}
\label{eq:variationLnDet}
\delta(\mbox{tr}\ln M[\phi])=\mbox{tr}(\delta M[\phi] M^{-1}[\phi])
\end{equation}
in order to determine the contribution of $\ln\det M[\phi]=\mbox{tr}\ln M[\phi]$
to the equations of motion~\refs{eq:hamEqs}. Since we only consider Yukawa couplings,
the variation $\delta M[\phi]$ enters~\refs{eq:variationLnDet} as
\begin{equation}
\label{eq:structurDeltaM}
\frac{\partial M[\phi]_{xy}}{\partial \phi_z} = 3g\phi_z\delta_{xz}\delta_{yz}.
\end{equation}
This holds true if the derivative of the superpotential appears in $M$ only on
the diagonal; this means that only a single matrix element of $M^{-1}[\phi]$ has
to be computed for each site $\phi_z$.\footnote{If this method was to be applied
to the Stratonovich prescription (where we can alternatively use~\refs{det11}),
two matrix elements would have to be computed since the interaction
is off-diagonal by one unit in the fermion matrix.} Obviously, we cannot avoid
an inversion of $M[\phi]$ in order to update all sites.

Finally let us briefly mention some details of our simulation runs. Since the
physical value of $m$ as well as the physical volume of the box $L$ were kept
fixed at $L=10\cdot m^{-1}=N a$, the lattice spacing $a$ was varied with the
number of points $N$ in the lattice. The lattice sizes we have considered
range from $N=15$ to $N=243$ corresponding to $a=0.06\ldots 0.004$. 
We have checked that the measurements were insensitive to finite size effects.
For all lattice sizes we generated between $250\,000$ and $400\,000$
independent configurations for the mass extraction; in order to
improve the signal to noise ratio for the Ward identities, $10^6$ independent
configurations were used (with $4$ independent runs in order to minimize
statistical errors).

\subsection{The two-dimensional Wess-Zumino model}
For the two-dimensional models, the fermion determinant is not
positive definite, in contrast to the situation in one dimension.
Indeed a numerical experiment shows that the determinant
has large fluctuations at strong coupling and changes its sign if $g/m$ is
larger than a certain threshold value of order $O(1)$. This value
depends on the lattice spacing (and on $m$). A closer look at the
spectrum of the fermion matrix reveals a second related problem,
namely the existence of very small eigenvalues. They increase the
condition number of the fermion matrix by several orders of magnitude
and prevent a straightforward application of the pseudofermion method.
In addition, the bosonic potential $|W'(\phi)|^2$ possesses two
separate minima; this might lead to complications with respect to
ergodicity. In order to check at which values of the coupling the
standard HMC breaks down, we have performed simulations in the weakly
coupled regime where the aforementioned contributions from the
fermionic fluctuations can be taken into account via reweighting
quenched ensembles. An added advantage is that with this approach,
field generations can be generated very quickly; on
the other hand, larger ensembles are used since reweighting requires
higher statistics.

\begin{figure}[!ht]
\begin{center}
  \includegraphics{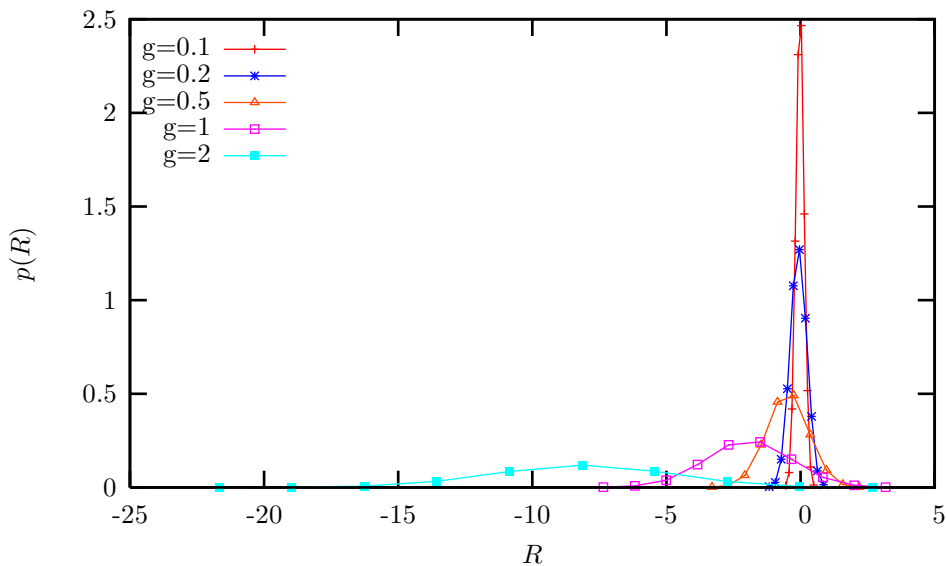}
  \caption{Probability distribution of $R=\ln\left(\det M/\det
  M_0\right)$. The stronger pronounced the peak the better the statistical
  errors are under control. The failure of the reweighting technique is
  visible for $g\geq 1$. The plotted data was generated from $20\,000$ configurations
for the model \emph{(iii)} on a $31\times 31$ square lattice.}
  \label{fig:reweight_comparison_fermdet}
\end{center}
\end{figure}

\begin{figure}[!ht]
\begin{center}
  \includegraphics{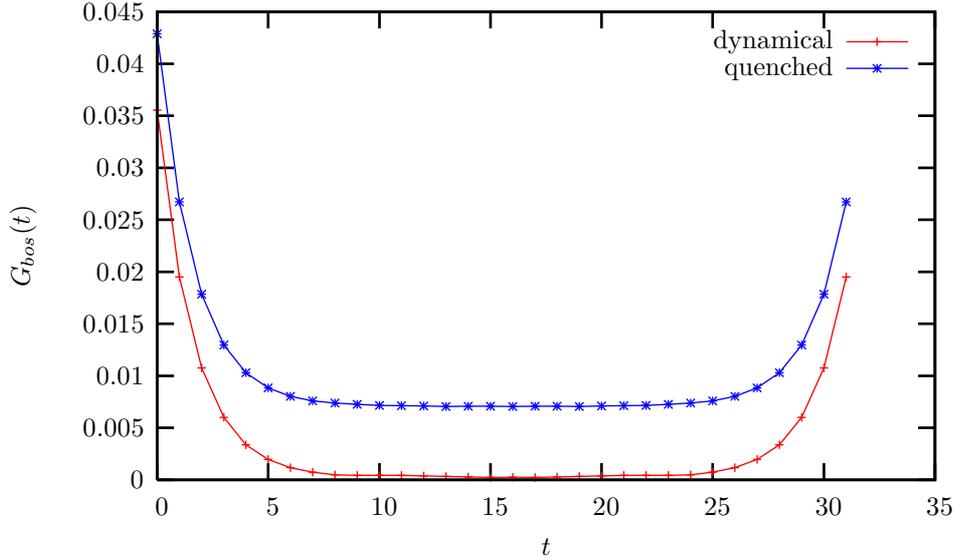}
  \caption{Comparison of the bosonic two-point function $G_{\rm bos}(t)$ for the
  quenched and full theory at $g=0.5$ for the model with standard Wilson term
  \emph{(i)} on a $32\times 16$ lattice. Similar results hold for the other
  models as well.}
  \label{fig:twopoint_comparison}
\end{center}
\end{figure}
Trial runs indicate that for the volumes considered the reweighting method
should work for $g\leq 1$. Figure~\ref{fig:reweight_comparison_fermdet}
shows that for these moderate couplings the fluctuations of the fermion determinant
are within two orders of magnitude. For $g=2$ the fluctuations span more than
ten orders of magnitude and the number of relevant configurations
becomes ridiculously small.

As pointed out in section~\ref{sec:WZmasswint}, for $g=0.5$
the perturbation of the mass is about 0.3\%, an effect which is clearly
not visible in our simulations. However, we can compare quenched with
reweighted expectation values and check whether they are sensitive to
the inclusion of dynamical fermions at all. We have found
that even in this weakly coupled regime the effective masses of bosonic and
fermionic superpartners coincide only if the fermion determinant is properly 
taken into account. Otherwise, the two-point functions (and hence the masses)
deviate considerably from their values in the full theory, as can bee seen in 
figure~\ref{fig:twopoint_comparison}.
Clearly, in order to simulate at larger couplings an improved algorithm
is needed. For example, a comparison with the results in~\cite{CatterallKaramov2}
would require a ratio of $g/m \simeq 0.3$, which
obviously is not feasible with the reweighting method used in this paper.

\subsection{Measurements and determination of masses}\label{sec:filter}
In this subsection we expand on the extraction of effective masses in our
quantum mechanical models. They are determined by the asymptotic behavior
of the (bosonic or fermionic) two-point functions $G(x)\simeq\sum_i
c_i e^{-(E_i-E_0)x}$:
For $x$ far away from the midpoint $N/2$ of the lattice, the asymptotic
expression for the bosonic two-point function is dominated by a single
mass state with $E_1-E_0=m_{\rm eff}$, and as a symmetric function it is proportional
to $\cosh(m_{\rm eff}(x-N/2))$; the fermionic two-point function as a superposition of
symmetric and antisymmetric parts (w.\,r.\,t.\ the midpoint)~%
\cite{CatterallKaramov2} is for small $x$ proportional to $e^{-m_{\rm eff}x}$. Thus,
the masses can be extracted by exponential fits from the corresponding
simulation data
\eqnl{
m_{\rm eff}\equiv
\log\left(\frac{G(x)}{G(x+a)}\right), \qquad G=G_{\rm bos}^{(n)}
\quad\mbox{or}\quad G_{\rm ferm}^{(n)}.}{massen3}
The $x$-region for the fit should be chosen in such a
way that
\begin{itemize}
  \item[(i)] the contribution of higher energy states with $E_i>E_1$ in the
    asymptotic expression $G(x)=\sum_i c_i e^{-(E_i-E_0)x}$ is negligible,
  \item[(ii)] the effect of the improvement terms, which violate reflection
    positivity (i.\,e., the $c_i$ are not necessarily positive) and
    therefore damp the two-point function for small values of $x$ below the
    continuum value, better not influence the result,
  \item[(iii)] the errors (which are larger for smaller values of the
    two-point function) should be minimized,
  \item[(iv)] and in the case of bosons, the influence from the second
    exponential tail in $\cosh(m_{\rm eff}(x-N/2))=
    \frac{1}{2}(e^{m_{\rm eff}(x-N/2)}+e^{-m_{\rm eff}(x-N/2)})$
    should not interfere with the exponential fit to the first.
\end{itemize}
Thus, we are supposed to choose a region for $x$ in between the left boundary
($x\gg 0$ by (i) and (ii)) and the midpoint ($x\ll N/2$ by (iii) and (iv)). This
works reasonably well for Wilson fermions (the results are presented in
section~\ref{sec:qmmasses}), but for SLAC fermions, we observe an oscillating
behavior of the two-point functions near the boundaries which requires a more
careful investigation.

\begin{figure}[!ht]
\begin{center}
  \includegraphics{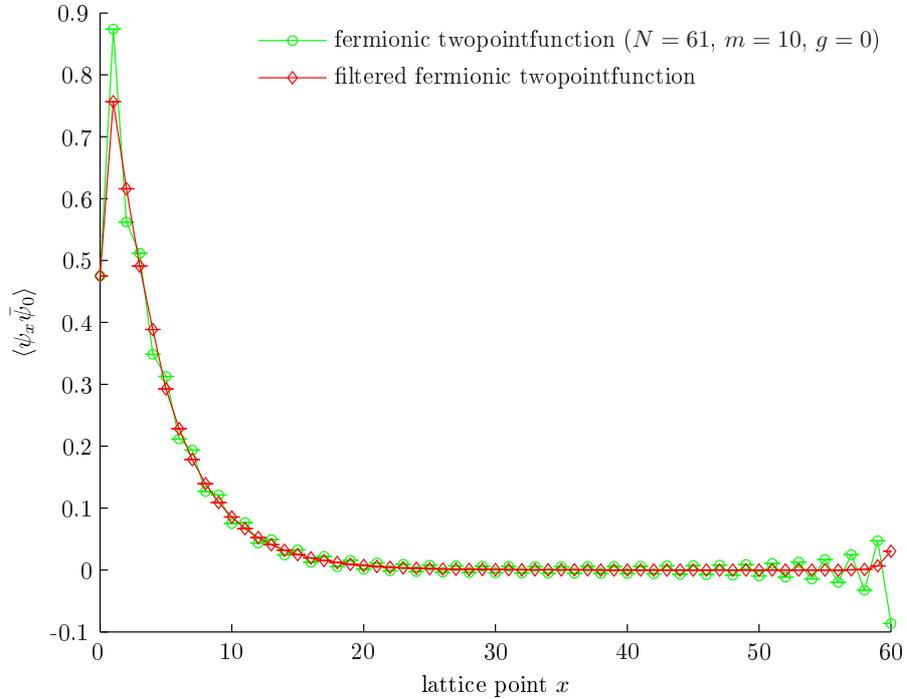}
  \caption{The free fermionic two-point function ($g_{\rm phys}=0$, $m_{\rm phys}=10$)
  before and after the application of the filter.}
  \label{fig:SLACMassesOscillate}
\end{center}
\end{figure}

In order to understand the nature of this phenomenon, we compute the propagator
of a free massive fermion in the continuum,
\eqnl{G_{\rm ferm}(x)=\sum_\l \frac{\psi_\l(x) \psi_\l^*(0)}{\l},}{slactwop1}
where the sum runs over the nonzero part of the spectrum of the differential
operator $\pa+m$, and $\psi_\l$ are the eigenfunctions of $\pa+m$ with
eigenvalue $\l$. If we impose periodic boundary conditions with
$\psi_\l(x)=\psi_\l(x+N)$, we have $\psi_\l(x)=\frac{1}{\sqrt{N}}
e^{\frac{2\pi i}{N}kx}$ with $\l=\frac{2\pi i}{N}k+m$ for all
integer $k$. Inserting this into~\refs{slactwop1}, we obtain
\eqnl{G_{\rm ferm}(x)=\frac{1}{N}\sum_k \frac{e^{\frac{2\pi i}{N}kx}}
  {\frac{2\pi i}{N}k+m}}{slactwop3}
for the two-point function on the circle. If we discretize the circle (and so
introduce a momentum cutoff), the sum in \refs{slactwop3} is truncated to a
finite number of terms and reduces just to the propagator of the SLAC derivative.
This cutoff in momentum space leads to the Gibbs phenomenon which is in fact what
we observe, e.\,g, in figure~\ref{fig:SLACMassesOscillate}.
\begin{figure}[!ht]
\begin{center}
  \includegraphics{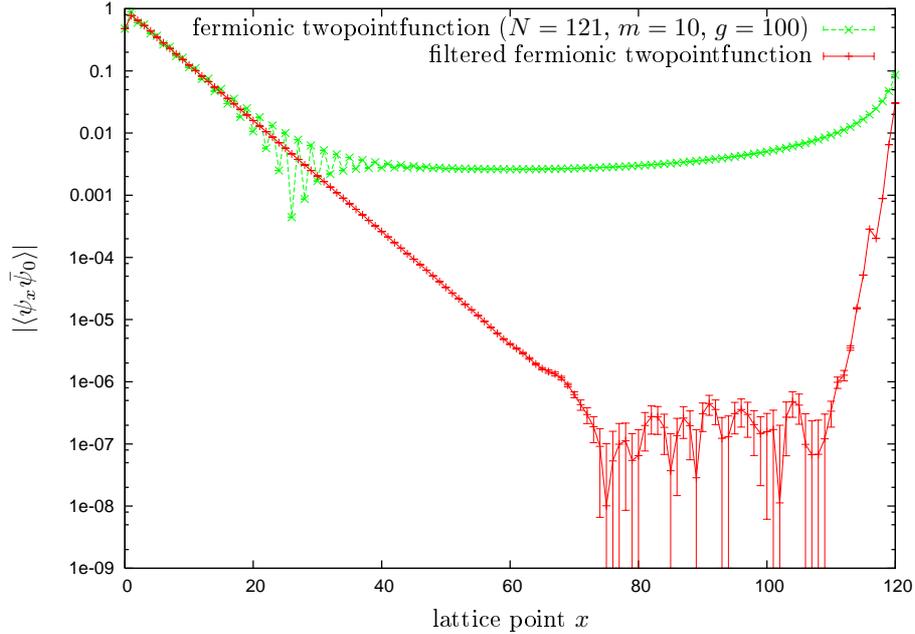}
  \caption{This plot shows one example of a fermionic two-point function of the interacting theory ($m_{\rm phys}=10$, $g_{\rm phys}=100$) in a logarithmic representation before and after the application of the filter. A linear fit of the filtered function yields the effective mass $m$. In the last part of the graph a large numerical error can be observed.}
  \label{fig:filterinteract}
\end{center}
\end{figure}

The error made by an approximation to a continuous and periodic function
by $n$ of its Fourier modes decreases exponentially with $n$. For a
discontinuous function, the error is proportional to some power
$n^{-\delta}$ away from the discontinuities. This can be improved by a filter
which increases the convergence rate~$\delta$ or even recovers the
exponential approximation of a continuous function.\footnote{In higher
dimensions, these Fourier approximation errors away from the discontinuities
are negligible in comparison to other errors.} For our computations, we have
used the ``optimal filter'' proposed in~\cite{filter}; the filter has compact
support in a region away from the discontinuities and is optimal in the sense
that it balances the competing errors caused by localizing it either in physical
or in momentum space. Its effect in the interacting case is illustrated in figure~\ref{fig:filterinteract}. In this logarithmic plot, it extends the
range from which one can extract the masses nearly up to the midpoint of the
lattice. The rather large deviations from the unfiltered data beyond that point
are irrelevant for our purposes.

For the two-dimensional models we have extracted the masses from the two point
correlators~\cite{CatterallKaramov2}
\begin{equation}
\label{eq:def_corellator_boson}
G_{\rm bos}(t)=\frac{1}{N_t N_x^2}\sum_{t'}\sum_{x,x'}\langle \varphi_2(t+t',x) \,\varphi_2(t',x')\rangle
\end{equation}
and 
\begin{equation}
\label{eq:def_corellator_fermion}
G_{\rm ferm}(t)=\frac{1}{N_tN_x^2}\sum_{\alpha=1}^2\sum_{t'}\sum_{x,x'}\langle \bar\psi_\alpha(t+t',x)
\,\psi_\alpha(t',x')\rangle
\end{equation}
with $N_t\equiv N_1$, $N_x\equiv N_2$. Taking spatial averages projects
onto $p_x\equiv p_2=0$; this explains the observed small discretization errors
for the choice $r^2=4/3$ in model~\emph{(ii)}, cf.\ section~\ref{sec:tdlatmod}.
Both correlation functions are proportional
to $\cosh(m(t-N_t/2))$. The mass can be extracted via a linear fit in the
logarithmic representation in analogy to the one-dimensional case.

%%%%%%%%%%%%%%%%%%%%%%%%%%%%%%%%%%%%%%%%%%%%%%%%%%%%%%%

\section{Renormalizibility of the WZ model with SLAC fermions}
\label{sec:three}
Kartens and Smit have demonstrated in~\cite{Karsten} that the SLAC
derivative induces a nonrenormalizable one-loop diagram in four-dimensional
quantum electrodynamics on the lattice.\footnote{In~\cite{Rabin}, it was
suggested that this might be resolved by a different resummation of the
perturbation series. But in the case of the Schwinger model, this prescription
does not have the correct continuum limit, as shown in~\cite{BodwinKovacs}.}
Therefore, theories involving the SLAC derivative were generally believed to be
nonrenormalizable. But it can be shown that the two-dimensional
${\cal N}=2$ Wess-Zumino model is in fact renormalizable at least to
one-loop order.\footnote{In fact, the discussion can easily be extended
to prove renormalizability also for the ${\cal N}=1$ model.}

As usual the calculations of lattice perturbation theory are carried out
in the thermodynamic limit where the number of lattice points tends to
infinity and the lattice momentum becomes continuous. In momentum space
the finite lattice spacing~$a$ is translated into a finite cutoff
$\Lambda=\frac{\pi}{a}$. It has to be shown that the diagrams can be
renormalized when this cutoff is removed. The BPHZ renormalization scheme
is used to prove that the renormalized integrals tend towards their continuum
counterparts, and the counterterms can be identified with similar quantities
of continuum perturbation theory. The argumentation employed here is closely
related to the renormalization theorem of Reisz \cite{Reisz}. This theorem
does however not apply in its original form because the integrands are not
smooth functions of the loop momentum.

In the following, we will determine an upper bound for the boson propagator
in momentum space which will be used later on to argue that parts of the
integrals in lattice perturbation theory are going to vanish in the continuum
limit. For the SLAC derivative, the momentum space representation of the
propagators
\begin{equation}
\frac{1}{P(k)^2+m^2}\quad\mbox{and}\quad
\frac{-i\slashed{P}(k)+m}{P(k)^2+m^2}
\end{equation} 
for bosons and fermions contains the saw tooth function
\begin{equation}
P_\mu(k)=k_\mu-2l\Lambda\quad\mbox{where}\quad (2l-1)\Lambda\leq k_\mu\leq
(2l+1)\Lambda .
\end{equation}
The momentum integration is always restricted to the first Brillouin zone,
$\mbox{BZ}=\{(k_\mu)|\,|k_\mu|\leq\Lambda\}$. The internal lines in
the one-loop diagrams carry either the internal momentum $k$ or a sum
$k+q$ of internal and external momenta where $q$ denotes a linear
combination of the external momenta (using momentum conservation, there
are $n-1$ such linear combinations $q_j$ in a diagram with $n$ vertices).
Integrations over loop momenta $k_\mu$ can be split into integrations over
a square $D=\{(k_\mu)|\,|k_\mu|\leq\frac{\pi\ve}{a}\}$ for an arbitrary
$\ve<\frac{1}{2}$ and the rest of the Brillouin zone, $\mbox{BZ}\backslash D$.
We will argue below that the integral over $D$ converges to the continuum
value of the integral whereas the integral over $\mbox{BZ}\backslash D$ is
shown to vanish as $a$ goes to zero.

Namely, for a given set of external momenta $\{q_j\}$, one may
choose $\eta=\max_{\mu,j}\{\frac{a_0}{\pi}|q_{j\mu}|\}$ with $a_0$ small
enough such that $0<\eta<\ve<\frac{1}{2}$. For $(k_\mu)\in D$, we
can then read off from
\eqnl{a |k_\mu \pm q_\mu|\leq a(|k_\mu| + |q_\mu|) < \pi(\ve + \eta)\quad
  \mbox{for all}\quad a<a_0}{eq:estimates1}
that $|k_\mu \pm q_\mu|\leq\Lambda\ve'$ with $\ve':=\ve + \eta < 1$, i.\,e.,
$(k_\mu \pm q_\mu)$ is also inside the first Brillouin zone. On the other
hand, if $(k_\mu)\in\mbox{BZ}\backslash D$,
\eqnl{\pi(\ve-\eta)\leq a(|k_\mu| - |q_\mu|)\leq a|k_\mu + q_\mu|\leq a(|k_\mu|
  + |q_\mu|)\leq \pi(1+\eta)}{eq:estimates2}
for such lattice spacings $a$. The latter inequality may be used in order
to find an upper bound for the propagator,
\eqnl{\frac{1}{P(k\pm q)^2 + m^2} < \frac{1}{P(k\pm q)^2} < C a^2}{renorm9}
with $C=\big((\ve-\eta)\sqrt{2}\pi\big)^{-2}$.

It can be easily seen that in the Wess-Zumino model, only two different types of
integrals contribute at one-loop level. A typical integral of the first type is
\begin{eqnarray}
  \label{eq:bosongeneral}
  I_\varepsilon +I_\pi & = & \int_{\mathrm{BZ}}\frac{d^2k}{(2\pi)^2}
    \frac{1}{(P(k)^2+m^2)(P(k+q_1)^2+m^2)\ldots(P(k+q_{n-1})^2+m^2)},\\
  I_\varepsilon & = & \int_D
    \frac{d^2k}{(2\pi)^2}\frac{1}{(k^2+m^2)((k+q_1)^2+m^2)\ldots((k+q_{n-1})^2+m^2)},\nonumber\\
  I_\pi & = & \int_{{\rm BZ}\backslash D}\frac{d^2k}{(2\pi)^2}
    \frac{1}{(k^2+m^2)(P(k+q_1)^2+m^2)\ldots(P(k+q_{n-1})^2+m^2)}\nonumber\\
  & \leq & (Ca^2)^{n-1}\int_{|k|\leq \sqrt{2}\pi/a}\frac{d^2k}{(2\pi)^2}\frac{1}{(k^2+m^2)}
    = (Ca^2)^{n-1} \log\Big(1+\frac{2\pi^2}{a^2 m^2}\Big).\nonumber
\end{eqnarray}
Here, we have applied~\refs{eq:estimates2} in order to find an upper bound for
the integrand in $I_\pi$ and then enlarged the integration domain to a full disk
including the first Brillouin zone. Thus, $I_\pi$ vanishes in the continuum limit
if $n>1$. Therefore, the integral $I_\varepsilon$ tends to the continuum value of
the integral as $a$ goes to zero (and the corresponding continuum integral is
convergent by power counting), so as long as we are considering diagrams with
more than one vertex, this type of integrals does not spoil
renormalizability.

Another class of integrals is
\begin{eqnarray}
  \label{eq:fermiongeneral}
  I'_\varepsilon +I'_\pi & = & \int_{\mathrm{BZ}}\frac{d^2k}{(2\pi)^2} \frac{P_\mu(k)
    P_\nu (k+\tilde{q}_1) \ldots P_\varrho(k+\tilde{q}_l)}{(P(k)^2+m^2)(P(k+q_1)^2+m^2)
    \ldots(P(k+q_{n-1})^2+m^2)},\\
  I'_\varepsilon & = & \int_D \frac{d^2k}{(2\pi)^2}\frac{k_\mu \ldots
    (k+\tilde{q}_l)_\varrho}{(k^2+m^2)((k+q_1)^2+m^2)\ldots((k+q_{n-1})^2+m^2)},\nonumber\\
  I'_\pi & = & \int_{{\rm BZ}\backslash D}\frac{d^2k}{(2\pi)^2}\frac{P_\mu(k)\ldots
    P_\varrho (k+\tilde{q}_l)}{(k^2+m^2)(P(k+q_1)^2+m^2)\ldots(P(k+q_{n-1})^2+m^2)}\nonumber\\
  & \leq & \int_{{\rm BZ}\backslash D} \frac{d^2k}{(2\pi)^2} \frac{|P_\mu(k)|\ldots
    |P_\varrho (k+\tilde{q}_l)|}{(k^2+m^2)(P(k+q_1)^2+m^2)\ldots(P(k+q_{n-1})^2+m^2)}\nonumber\\
  & \leq & \left(\frac{\pi}{a}\right)^{l+1}(Ca^2)^{n-1}\int_{|k|\leq \sqrt{2}\pi/a}
    \frac{d^2k}{(2\pi)^2}\frac{1}{(k^2+m^2)}
    = C^{n-1}a^{2n-l-3} \log\Big(1+\frac{2\pi^2}{a^2 m^2}\Big).\nonumber
\end{eqnarray}
The $\tilde{q}_i$ are taken from the $q_j$, so $l\leq n-1$. The same arguments as above
show that the continuum limit is correct for any $n>2$ (again, all continuum integrals
are convergent by power counting).

Therefore, renormalizability only remains to be shown for two kinds of integrals.
The first consists of diagrams with $n=1$, e.\,g., tadpole diagrams. In this case,
the loop momentum is independent of the (vanishing) exterior momentum so that the
argument of $P_\mu(k)$ is restricted to the first Brillouin zone (where
$P_\mu(k)=k_\mu$). The boundary of the integration region behaves as a finite cutoff
that is removed in the continuum limit so that the integral approaches its continuum
counterpart. In the BPHZ renormalization scheme these diagrams are just subtracted
and do not contribute to the renormalized quantities.

The second kind of integrals (with $n=2$ and $l=1$ in \refs{eq:fermiongeneral})
requires a more careful investigation which may be found in appendix~\ref{sec:AppA}.
This demontrates that lattice discretizations of the two-dimensional ${\cal N}=2$
Wess-Zumino model based on the SLAC derivative are renormalizable at first order in
perturbation theory and yield the correct continuum limit. It seems however
problematic to use the BPHZ renormalization scheme to renormalize the
corresponding diagrams in higher-dimensional cases since this would require a
differentiation of the integrands with respect to external momenta. The discontinuity
of the saw tooth functions in this case would lead to singular terms.

\section{Summary and outlook}
In this paper, we have studied supersymmetric ${\cal N}=2$ Wess-Zumino models
in one and two dimensions. The six quantum mechanical models under consideration
differ by the choice of lattice derivatives and improvement terms. The latter
can be used to render the theory manifestly supersymmetric on the lattice;
in distinction to previous works on this subject, our simulations of the
broken Ward identities at strong coupling prove that only one supersymmetry
can be preserved. We have demonstrated to a high numerical precision that by
far the best results for bosonic and fermionic masses can be obtained from a
model with Wilson fermions and Stratonovich prescription for the evaluaton of
the improvement term and from a model based on the SLAC derivative. It is
interesting to note that for SLAC fermions no improvement term is needed to
recover supersymmetry in the continuum limit.

As a key result of this paper for two-dimensional Wess-Zumino models, we propose
a non-standard Wilson term giving rise to an $O(a^2)$ improved Dirac operator
\eqnl{
M=\gamma^\mu\ring{\pa}_\mu +\frac{iar}{2}\gamma_3\Delta\mtxt{with} r^2=\frac{4}{3}.}{conc1}
The masses extracted from this model approach the continuum values much faster
than those for the model with standard Wilson-Dirac operator
\refs{stand3}. Again, results of a comparably good quality can be obtained with
nonlocal SLAC fermions. In our case, the common
reservation that the SLAC derivative leads to non-renormalizable theories
(as originally shown in~\cite{Karsten} for the case of four-dimensional gauge
theory) does not hold; we have proven that the Wess-Zumino model in two
dimensions with this derivative is renormalizable to one-loop order.

Motivated by the fact that the masses for the ${\cal N}=2$ Wess-Zumino
lattice model with ultralocal Dirac operator~\refs{conc1} are quite close
to the continuum values already for moderate lattices we plan to study
the model at strong couplings where we will see deviations from the free
theory. We are about to implement the PHMC algorithm~\cite{Frezzotti:1997ym}
as a possibility to deal with the small eigenvalues of the fermionic operator.
We believe that the ${\cal N}=2$ Wess-Zumino model as a simple and
well-understood theory without the complications of gauge fields has the
potential to become a toy-model for developing efficient algorithms for
systems with dynamical fermions, similar to the ubiquitous Schwinger
model which serves as toy model for more complex systems with a chiral
condensate, instantons, confinement and so forth.

Wess-Zumino models are the flat-space limits of Landau-Ginzburg models.
A related project might be to study another limit of Landau-Ginzburg theories,
namely sigma models in non-trivial K\"ahlerian without a superpotential.
Such sigma models admit two supersymmetries; typically they have
instanton solutions and chiral condensates may be generated. If there exist
local Nicolai variables which give rise to improved lattice models
with one quarter of supersymmetry, contact could be made with our
investigations of Wess-Zumino models in a much broader physical context.
Clearly these interesting field theories deserve further attention, both from
the algebraic and from the numerical side.

A further obvious problem is to study the nonperturbative sector of the 
two-dimensional ${\cal N}=1$ Wess-Zumino model. This model shows a richer phase
structure than the model with two supersymmetries. The sign problem for the
Pfaffian seems unavoidable. It is interesting to note that in conventions
with hermitean gamma matrices, a nonvanishing Wilson term for Majorana fermions
has to enter the Dirac operator as in~\refs{conc1}. Nevertheless, the discretization
errors in this case will be of order $O(a)$; this can in principle be improved
to $O(a^2)$ by a slightly less natural substitute for the Wilson term. In general,
for the ${\cal N}=1$ model, no local Nicolai variables can be constructed which
would suggest a supersymmetric completion of the naive lattice action. However,
one might expect (as in quantum mechanics) that such improvement terms for the
two-dimensional ${\cal N}=1$ model with SLAC fermions are in fact dispensable.

\label{sec:SO}
\subsection*{Acknowledgements}
We thank F.~Bruckmann, S.~Catterall, S.~Duerr, C.-P.~Georg, J.~Giedt, K.~Jansen
and C.~Wozar for discussions. T.~Kaestner acknowledges support by the
Konrad-Adenauer-Stiftung and G.~Bergner by the Evangelisches Studienwerk.
This work has been supported by the DFG grant Wi 777/8-2.

\begin{appendix}
\section{Renormalization of the fermion loop}
\label{sec:AppA}
\Revision{$Rev: 903 $}
In section~\ref{sec:three}, we have demonstrated which kinds of integrals are
potentially dangerous for the one-loop renormalizability of the ${\cal N}=2$
Wess-Zumino model. The missing integral in this proof was given by~%
\refs{eq:fermiongeneral} with $n=2$ and $l=1$; it appears for
fermion loops with two internal lines,
\begin{equation}
  \label{eq:nontrivdiag}
  \raisebox{\shiftfeyn}{\includegraphics{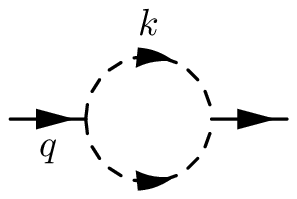}}\:\:\propto\:\:\int_{\mathrm{BZ}}
  \frac{d^2 k}{(2\pi)^2}\frac{P_\mu(k)P^\mu(k-q)}{(P^2(k)+m^2)(P^2(k-q)+m^2)}.
\end{equation}
Depending on the superpotential more than two external lines may appear;
in this case $q$ denotes the sum of all incoming external momenta. In order
to facilitate the evaluation of this integral, we add (and subtract) a term
with a finite continuum limit; apart from that, the integral needs to be
regulated in this limit. On the lattice with finite lattice spacing, BPHZ
regularization means to subtract the (as yet finite) value of the integral
with vanishing exterior momenta. Thus, we consider
\begin{eqnarray}
  & & \int_{\mathrm{BZ}}\frac{d^2 k}{(2\pi)^2}\frac{P_\mu(k)P^\mu(k-q)}{(P^2(k)+m^2)(P^2(k-q)+m^2)}
    + \int_{\mathrm{BZ}} \frac{d^2k}{(2\pi)^2}\frac{m^2}{(P^2(k)+m^2)(P^2(k-q)+m^2)}\nonumber \\
  & & {}- (\mbox{value at }q=0)\nonumber \\
  & = & \int_{\mathrm{BZ}}\frac{d^2 k}{(2\pi)^2}
    \frac{P^\mu(k-q)\big(k_\mu-P_\mu(k-q)\big)}{(P^2(k)+m^2)(P^2(k-q)+m^2)}\nonumber \\
  & = & \int_{\rm BZ}\frac{d^2 k}{(2\pi)^2}\frac{q^\mu P_\mu(k-q)}{(k^2+m^2)(P(k-q)^2+m^2)}\nonumber \\
  & & \qquad\qquad\qquad\qquad{}- 2\Lambda\sum_\mu\int_{-\Lambda}^{-\Lambda+q_\mu}\!\!\!\! dk_\mu
    \int_{-\Lambda}^\Lambda \frac{dk_{\nu\neq\mu}}{(2\pi)^2}\frac{P^\mu(k-q)}{(k^2+m^2)(P(k-q)^2+m^2)}.
    \label{loop1}
\end{eqnarray}
In the last step, we have chosen $a_0$ in such a way that for all $a=\pi/\Lambda<a_0$, shifting
$k_\mu\in\mbox{ BZ}$ by $-q_\mu$, one winds up either in the same or in an adjacent Brillouin zone, i.\,e.,
\eqnl{P_\mu(k-q)=
  k_\mu-q_\mu+2\Lambda\big(\Theta(-\Lambda-k_\mu+q_\mu)-\Theta(k_\mu-q_\mu-\Lambda)\big).}{loop3}
The first term on the right-hand side of~\refs{loop1} can be easily seen to
converge to the value of its continuum counterpart by similar arguments as
in~\refs{eq:bosongeneral} and \refs{eq:fermiongeneral}. In order to prove that
the second term does not give rise to any corrections in the continuum limit,
we make use of~\refs{eq:estimates1} and~\refs{renorm9} and observe that an
upper bound for its modulus is given by
\begin{eqnarray}
  & & 2\Lambda\sum_\mu\int_{-\Lambda}^{-\Lambda+q_\mu}\!\!\!\! dk_\mu
    \int_{-\Lambda}^\Lambda \frac{dk_{\nu\neq\mu}}{(2\pi)^2}\frac{|P^\mu(k-q)|}{(k^2+m^2)
    (P(k-q)^2+m^2)}\nonumber\\
  & \leq & 2C \pi^2\int_{-\Lambda}^{-\Lambda+q_1}\frac{d k_1}{2\pi}\int_{-\Lambda}^{\Lambda}
    \frac{d k_2}{2\pi}\frac{1}{k^2+m^2} + (q_1\leftrightarrow q_2, k_1\leftrightarrow k_2)
    \nonumber \\
  & = & \frac{C}{2}\left|\int_{-\Lambda}^{\Lambda}d k_2 \arctan\Big(
    \frac{q_1}{\omega(k_2)-q_1\Lambda \omega(k_2)^{-1} + \Lambda^2 \omega(k_2)^{-1}}\Big)
    \omega(k_2)^{-1}\right| + (q_1\leftrightarrow q_2, k_1\leftrightarrow k_2)\nonumber\\
  & \leq & \frac{C}{2}\int_{-\Lambda}^{\Lambda} d k_2
    \Big|\frac{q_1}{m^2+k_2^2-\Lambda q_1+\Lambda^2}\Big| + (q_1\leftrightarrow q_2,
    k_1\leftrightarrow k_2)
\end{eqnarray}
with $\omega(k) = \sqrt{m^2+k^2}$; here, we have also used that $|\arctan(x)|\leq |x|$.
It is obvious that this upper bound converges to zero in the limit where the lattice
cutoff is removed.

This completes the proof that the discretization of the ${\cal N}=2$ Wess-Zumino model
based on the SLAC derivative is one-loop renormalizable.
\end{appendix}

\end{document}